\documentclass{emulateapj}

\newcommand{\etal}{{et al.\ }}
\newcommand{\be}{\begin{equation}}
\newcommand{\ee}{\end{equation}}
\newcommand{\bea}{\begin{eqnarray}}
\newcommand{\eea}{\end{eqnarray}}
\newcommand{\erf}{\mathrm{erf}}

\shorttitle{Eccentric Extrasolar Planets}
\shortauthors{Ford and Rasio}
\slugcomment{to accepted in ApJ}

\begin{document}

\title{Origins of Eccentric Extrasolar Planets:  
       Testing the Planet--Planet Scattering Model}

\author{Eric B.\ Ford\altaffilmark{1,2,3} and 
        Frederic A.\ Rasio\altaffilmark{4}}

\email{eford@astro.ufl.edu}
\altaffiltext{1}{Department of Astronomy, University of Florida, 211 Bryant Space Science Center, P.O. Box 112055, Gainesville, FL 32611-2055}
\altaffiltext{2}{Hubble Fellow}
\altaffiltext{3}{Harvard-Smithsonian Center for Astrophysics, Mail Stop 51, 
                  60 Garden Street, Cambridge, MA 02138}
\altaffiltext{4}{Department of Physics and Astronomy, Northwestern University, 
                 Evanston, IL 60208}

\begin{abstract}
In planetary systems with two or more giant planets, dynamical
instabilities can lead to collisions or ejections through strong
planet--planet scattering. Previous studies for simple initial
configurations with two equal-mass planets revealed two discrepancies
between the results of numerical simulations and the observed orbital
elements of extrasolar planets: the potential for frequent collisions
between giant planets and a narrow distribution of final
eccentricities following ejections.  Here, we show that simulations
with two {\em unequal mass\/} planets starting on nearly circular
orbits predict a reduced frequency of collisions and a broader range
of final eccentricities.  We show that the two-planet scattering model
can easily reproduce the observed eccentricities with a plausible
distribution of planet mass ratios.  Further, the two-planet
scattering model predicts a maximum eccentricity of $\simeq0.8$,
independent of the distribution of planet mass ratios, provided that
both planets are initially place on nearly circular orbits.  This
compares favorably with current observations and will be tested by
future planet discoveries.  Moreover, we show that the combination of
planet--planet scattering and tidal circularization may be able to
explain the existence of some giant planets with very short period
orbits.  Orbital migration due to planet scattering could play an
important role in explaining the increased rate of giant planets with
orbital periods of less than a year, as found by radial velocity
surveys.  We also re-examine and discuss various possible correlations
between eccentricities and other properties of observed extrasolar
planets.  We find that the radial velocity observations are consistent
with planet eccentricities being correlated with the ratio of the
escape velocity from the planet's surface relative to the escape
velocity from the host star at the planet's location.  We demonstrate
that the observed distribution of planet masses, orbital periods, and
eccentricities can provide constraints for models of planet formation
and evolution.
\end{abstract}

\section{Introduction} 
\label{sec_intro}

For several centuries, theories of planet formation had been designed
to explain our own Solar System, but the first few discoveries of
extrasolar planets immediately sent theorists back to the drawing
board.  These discoveries led to the realization that planet formation
theory must be generalized to explain a much wider range of properties
for planetary systems.  For example, it had long been assumed that
planets formed in circular orbits because of strong eccentricity
damping in the protoplanetary disk and that their orbits would later
remain nearly circular (i.e., with eccentricity $e\le$0.1; Lissauer
1993, 1995).  However, over half of the extrasolar planets beyond
$0.1\,$AU have eccentricities $e\ge$0.3, and two have eccentricities
larger than 0.9.

The planets in eccentric orbits are generally believed to have formed
on nearly circular orbits but later evolved to their presently
observed large eccentricities.  Theorists have suggested numerous
mechanisms to excite the orbital eccentricity of giant planets.  These
include:
\begin{itemize}
\item[a)] secular perturbations due to a distant stellar or massive
planetary companion (Holman, Touma, \& Tremaine 1997; Mazeh et al.\
1997; Ford, Kozinsky, \& Rasio 2000; Takeda \& Rasio 2005),
\item[b)] perturbations from passing stars (Laughlin \& Adams
1998; Hurley \& Shara 2002; Zakamska \& Tremaine 2004),
\item[c)] strong planet--planet scattering events in planetary systems
with either a few planets (Rasio \& Ford 1996; Weidenschilling \&
Marzari 1996; Ford, Havlickova, \& Rasio 2001 (FHR); Marzari \&
Weidenschilling 2002; Yu \& Tremaine 2001; Ford, Rasio \& Yu 2003;
Veras \& Armitage 2004, 2005, 2006) or many planets (Lin \& Ida 1997;
Levison et al.\ 1998; Papaloizou \& Terquem 2001; Adams \& Laughlin
2003; Moorhead \& Adams 2005; Goldreich, Lithwick, \& Sari 2004; Ford \& Chiang 2007; Juric \&
Tremaine 2008),
\item[d)] interactions of orbital migration with mean-motion
resonances (Chiang \& Murray 2002; Kley 2000; Kley et al.\ 2004, 2005;
Lee \& Peale 2002; Tsiganis et al.\ 2005),
\item[e)] resonances between secular perturbations and precession
induced by general relativity, stellar oblateness, and/or a remnant
disk (Ford et al.\ 2000; Nagasawa et al.\ 2003; Adams \&
Laughlin 2006), 
\item[f)] interactions with a planetesimal disk (Murrary et al. 1998),
\item[g)] interactions with a gaseous proto-planetary disk (Goldreich \&
Tremaine 1980; Artymowicz 1992; Papaloizou \& Larwood 2000; Papaloizou et al.\ 2001; Goldreich \& Sari 2003; Ogilvie \& Lubow 2003; Cresswell et al.\ 2007; Moorhead \& Adams 2008),
\item[h)] asymmetric stellar jets (Namouni 2005, 2006), and
\item[i)] hybrid scenarios that combine aspects of more than one of
the above mechanisms (e.g., Marzari et al.\ 2005; Sandor \& Kley 2006;
Malmberg et al.\ 2007ab).
\end{itemize}
Some of mechanisms (a, b) inevitably influence the evolution of some
planetary systems, but are not able to explain the ubiquity of
eccentric giant planets (Zakamska \& Tremaine 2004; Takeda \&
Rasio 2005).  Observations of multiple planet systems have provided
strong evidence that other mechanisms (c, d) are also significant in
altering planet's orbital eccentricities.
For example, the dramatic eccentricity oscillations of $\upsilon$ And
c provide an upper limit on the timescale for eccentricity excitation
in $\upsilon$ And ($\simeq 100$yr) and strong evidence for
planet--planet scattering in this system (Ford, Lystad \& Rasio 2005).
Other multiple planet systems may also exhibit similar behavior
(Barnes \& Greenberg 2006ab).  Simulations of possible progenitors to
our own outer solar system have shown that instabilities can be
postponed while there is a significant disk mass and become manifest once the mass
of the disk decreases (relative to the planets' mass).  Further, the
remaining disk need not be sufficiently massive to damp the
eccentricities of eccentric planets that emerge from the instability
(Ford \& Chiang 2007; Chatterjee et al.\ 2008).
As another example, the detection of pairs of planets in 2:1 mean
motion resonances (e.g., GJ 876 b \& c) suggests that smooth
convergent migration likely occurred in these systems.  Additionally,
the fact that migration models can simultaneously match the observed
eccentricities for both planets b \& c suggests eccentricity
excitation was related to the migration and resonant capture in this
system (Lee \& Peale 2002; Kley et al.\ 2005).  
It is not clear if the remaining mechanisms (e-h) are
important for shaping the actual distribution of planet
eccentricities.

In this paper, we expand upon the original planet--planet scattering
model of Rasio \& Ford (1996) and FHR.  First, we evaluate some
potential origins of dynamical instabilities that result in close
encounters and strong planet--planet scattering in \S2.  In \S3, we
present the results of n-body simulations of planet--planet scattering
for systems with two giant planets of unequal masses.  Then, in \S4,
we compare the predictions of eccentricity excitation models with the
eccentricities of the known extrasolar planets.  In \S5, we discuss
the implications of our work for theories of eccentricity excitation
and damping and suggest how future observations can further test
theories for eccentricity excitation.

\section{Origin of Instability}
\label{sec_instability}

While some authors have simulated multiple planet systems beginning
with the planet formation stage, computational cost has
limited such simulations to a small portion of the disk and/or small
number of initial conditions
(e.g., Kokubo \& Ida 1998; Levison, Lissauer, \& Duncan 1998).  Since
dynamically unstable planetary systems are highly chaotic, we can only
investigate the statistical properties of an ensemble of 
systems with similar initial conditions.  Thus, most investigations of
dynamical instabilities in multiple planet systems proceed by
simulating systems after planets have formed and
perturbations due to the protoplanetary disk are no longer
significant.  The planets are placed on plausible initial orbits and
numerically integrated according to the gravitational potential of the
central star and other planets.

Clearly, the choice of initial conditions will determine whether the
systems are dynamically stable and will affect the outcome of unstable
systems.  Our simulations of planet--planet scattering typically begin
with closely spaced giant planets (e.g., Rasio \& Ford 1996; FHR).
This is necessary for dynamical instabilities to occur in systems with
only two planets initially on circular orbits.  For two-planet
systems, there is a sharp transition from rigorous Hill stability to
chaos and strong interactions.  Therefore, one potential concern about
the relevance of dynamical instabilities is whether the necessary
initial conditions will manifest themselves in the two-planet
configurations that occur in nature.  In this section, we describe
several possible mechanisms that could lead to dynamical instabilities
in two-planet systems, including mass growth through accretion,
dissipation of the protoplanetary disk, and orbital migration.
Additionally, the secular evolution of systems with more than two
planets provides a natural mechanism for triggering dynamical
instabilities, even long after the protoplanetary disk has dissipated
and planets are fully assembled (Chatterjee et al.\ 2008).

According to the standard core accretion model, once a rocky planetary
core reaches a critical mass, it rapidly accretes the gas within its
radius of influence in a circumstellar disk.  Thus, the semi-major
axis of a planetary core is determined by the collisional evolution of
protoplanets, while the mass of a giant planet is determined by the
state of the gaseous disk when the core reaches the critical mass
(Lissauer 1993).  Two planetary cores could form with an initial
separation sufficient to prevent close encounters while their masses
are less than the critical mass for runaway accretion, but
insufficient to prevent a dynamical instability after the onset of
rapid mass growth due to gas accretion (Pollack et al.\ 1996).  For
closely packed massive giant planets, the final stage of mass growth
may occur via accretion through a common gap (e.g., Schafer et al.\
2004; Sandor et al.\ 2007).

The accumulation of random velocities provides another possible source
of a dynamical instability.  Assuming planets form in the presence of
a dissipative disk, they are expected to form on nearly circular and
coplanar orbits.  While the timescale for dissipation in the disk
remains shorter than the timescales for eccentricity excitation,
eccentricities and inclinations will be damped, preventing close
encounters.  Both analytical and numerical studies of eccentricity
damping in gaseous disks suggest that the eccentricity damping
timescale is much shorter than the migration timescale (Goldreich \&
Tremaine 1980; Trilling et al.\ 1998; Nelson et al.\ 2000; Papaloizou
et al.\ 2001).  According to a dynamical analyses of the GJ 876
multiple planet systems, the current eccentricities suggests that
eccentricity damping timescale must have been at least 40 times more
rapid than the migration timescale (Lee \& Peale 2002; Kley et al.\
2005).  Thus, planets are assumed to remain on nearly circular orbits
during putative early migration stage.  As the disk dissipates,
eccentricity damping becomes less significant, so mutual planetary
perturbations can excite significant eccentricities and inclinations
and lead to close encounters between planets.  Since the
photoevaporation timescale ($\sim10^5$ yr; Alexander et al. 2006) is
often much shorter than the timescale for dynamical instabilities to
arise, the outcome of the instabilities are expected to be insensitive
to the details of the photoevaporation process.  We have begun to
investigate the dynamical evolution of multi-planet systems that are
interacting with a gas disk in order to test this assertion and to
better understand the implications of trapping in mean motion
resonances for the onset of dynamical instabilities (Chatterjee et al.\ 2008; Payne et al.\ in prep).  The results of
such simulations are beyond the scope of this paper and will be
reported in future papers.

Finally, the discovery of giant planets at small orbital separations
suggests that orbital migration may be common.  In multiple planet
systems, convergent migration (i.e., with the ratio of semi-major axes
approaching unity) could increase the strength of mutual planetary
perturbations and excite eccentricities (even before/without resonant
capture).  For systems with exactly two giant planets, then the
stability boundary (assuming nearly circular orbits) occurs inside the
2:1 mean motion resonance.  Therefore, if systems form with a ratio of
orbital periods exceeding 2:1, then a smooth migration would be
expected to result in capture into the 2:1 mean motion resonance.
Continued migration is expected to result in significant eccentricity
excitation (e.g., Lee \& Peale 2002).  Indeed, this can lead to the
onset of dynamical instabilities and strong planet-planet scattering
(e.g., Sandor \& Kley 2006; Sandor et al.\ 2007).  Similar outcomes
may occur due to trapping in the other mean motion resonances (e.g.,
3:1, Adams \& Laughlin 2003; 3:2 or perhaps 5:3, Lee et al.\ 2008).  Our
choice of initial conditions in this paper is not intended to
represent this scenario.  We intend to explore this scenario in future
investigations.

In contrast to the case of two-planet systems, there is no sharp
stability criterion for three-planet systems.  Three-planet systems
can be unstable even for initial orbital spacings significantly
greater than would be necessary for similar two-planet systems to be
unstable (Chambers, Wetherill \& Boss 1996).  Additionally, such
systems can evolve quasi-stably for very long times,
$\sim10^{6}-10^{10}\,$yr, before chaos finally leads to close
encounters and strong planet--planet scattering (Marzari \&
Weidenschilling 2002; Chatterjee at al.\ 2008).  This longer timescale
until close encounters could allow sufficient time for three or more
planets to form via either the disk instability or core accretion
models.

If protoplanetary disks form many planets nearly simultaneously, then
planet--planet scattering may lead to a phase of dynamical relaxation.
Several researchers have numerically investigated the dynamics of
planetary systems with $\sim10-100$ planets (Lin \& Ida 1997; Levison
et al.\ 1998; Papaloizou \& Terquem 2001, 2002; Adams \& Laughlin
2003; Barnes \& Quinn 2004; Juric \& Tremaine 2008; Payne et al.\ in
prep).  Initially, such systems are highly chaotic and close
encounters are common.  The close encounters lead to planets colliding
(creating a more massive planet) and/or planets being ejected from the
system, depending on the orbital periods and planet radii.  Either
process results in the number of planets in the system being reduced
and the typical separations between planets increasing.  The system
gradually evolves from a highly unstable state to quieter states,
which can last longer before the next collision or ejection.  Such
systems typically evolve ultimately to a final state with 1--3
eccentric giant planets that will persist for the lifetime of the star
(Adams \& Laughlin 2003; Juric \& Tremaine 2008).

With so many possibilities for triggering dynamical instabilities in
multiple planet systems, we expect that these processes may be rather
ubiquitous.  While real planetary systems likely have more than two
massive bodies, simulations of relatively simple systems (e.g., with
just two giant planets) facilitate the systematic study of the
relevant physics and help develop intuition for thinking about the
evolution of more complex systems.

\section{Numerical Investigation of Planet--Planet Scattering}
\label{sec_numint}

In the previous section, we argued that if planet formation commonly
results in planetary systems with multiple planets, then it should be
expected that the initial configurations will not be dynamically
stable for time spans orders of magnitude longer than the timescale
for planet formation.  Shortly after the discovery of the first eccentric
extrasolar planets, Rasio \& Ford (1996) conducted Monte Carlo
integrations of planetary systems containing two equal-mass giant planets
initially placed just inside the Hill stability limit (Gladman 1993).
They numerically integrated the orbits of such systems until there was
a collision, or one planet was ejected from the system, or some
maximum integration time was reached.  The two most common outcomes
were collisions between the two planets, producing a more massive
planet in a nearly circular orbit between the two initial orbits, and
ejections of one planet from the system while the other planet remains
in a tighter orbit with a large eccentricity.  The relative frequency
of these two outcomes depends on the ratio of the escape velocity from the surface of the planet to the escape velocity from the host stars at the planet's location (see \S4.3.3).

While this model could naturally explain how planets acquire large
eccentricities, FHR performed a large ensemble of planet--planet
scattering experiments to compare the resulting planetary systems to
the observed sample and found two important differences.  First, for
the relevant radii and semi-major axes, collisions of Jupiter-mass
planets were more frequent in the simulations than nearly circular
orbits are observed among the known extrasolar planets.

However, the branching ratios from those simulations may not be
appropriate for realistic planetary systems.  Since there is a sharp
and rigorous Hill stability limit for two-planet systems, the initial
conditions placed the two planets in orbits with a relatively small
separation.  Since FHR also assigned the planets small initial
eccentricities and inclinations, the planets initially had a small
relative velocity at conjunction (compared to their circular velocity)
and gravitational focusing increased the rate of collisions early on
in the simulations.  The rate of collisions drops significantly (for
the systems that survive long enough) once the planets have had time
to excite each other's eccentricities.  Thus, the fraction of systems
that result in collisions is likely sensitive to the initial
conditions.  

To determine more accurately the fraction of actual two-planet systems
that result in collisions, future studies would need
to model the onset of the instability more realistically.
Unfortunately, direct n-body integrations of young planetary systems
with small bodies are extremely computationally demanding.  The
significance of initial conditions is less pronounced for n-body
integrations of systems with three or more planets, since more distant
initial spacings can be used, so that all close encounters occur only
after the planets have excited each other's eccentricities.  Despite
these potential complications, it can be useful to study relatively
simple model systems to develop intuition for more complex problems
and to understand the limitations of simple models.  In that spirit,
FHR reported the results of planet--planet scattering experiments
involving two equal-mass planets, while here we report the results of
planet--planet scattering experiments involving two planets of unequal
masses.  

\begin{figure}
\epsscale{1.0}
\plotone{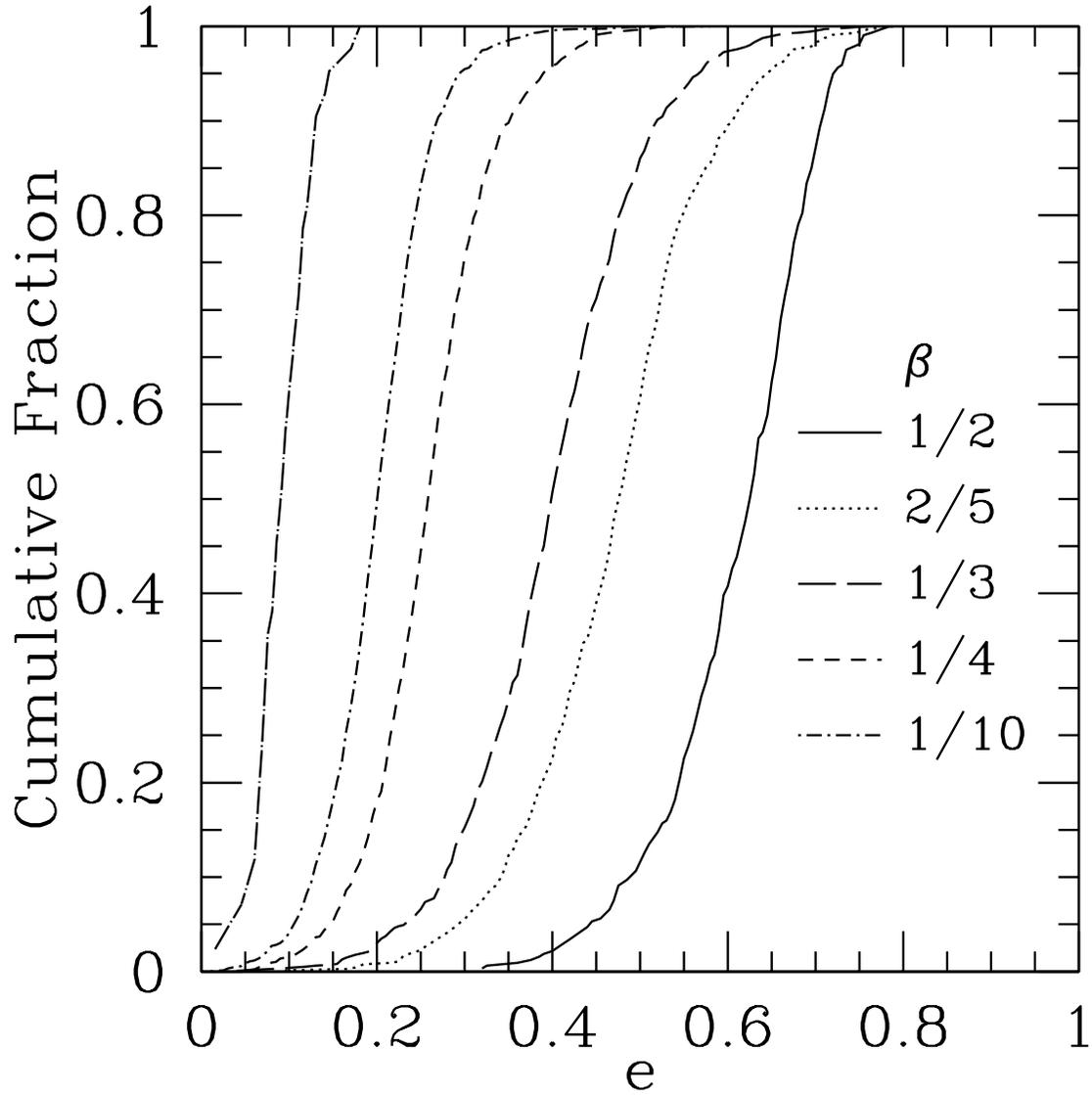} 
\caption{\label{fig_evsbeta}
Cumulative distributions for the final eccentricity of
the remaining planet after the other planet has been ejected.  Each
line style corresponds to an ensemble of simulations with
$\mu_1+\mu_2=6\times10^{-3}$ and a different value of
$\beta\equiv~m_f/(m_1+m_2)$, where $m_f$ is the mass of the 
mass of the putative ejected planet.
}
\end{figure}

The more significant shortcoming of the two equal-mass planet
scattering model identified by FHR was that, in systems leading to
one ejection, the eccentricity distribution of the remaining planet
was concentrated in a narrow range and was greater than the typical
eccentricity of the known extrasolar planets (See Fig.~\ref{fig_evsbeta},
right, rightmost curve).  FHR speculated that planet--planet scattering
involving two planets of unequal masses would result in planets
remaining with a broader distribution of eccentricities.  In this
section, we present results that confirm this speculation and quantify
the resulting eccentricities.  

\subsection{Initial Conditions}

We used the mixed variable symplectic algorithm of Wisdom
\& Holman (1991), modified to allow for close encounters between planets as
implemented in the publicly available code {\tt Mercury} (Chambers 1999).
The results presented below are based on $\sim 10^4$ numerical
integrations.  
Our numerical integrations were performed for a system containing two
planets, with mass ratios $10^{-4} < \mu_i < 10^{-2}$, where
$\mu_i\equiv~m_i/M_\star$, $m_i$ is the mass of the $i$th planet, and
$M_\star$ is the mass of the central star.  We use $i=1$ to denote the
planet initially closer to the star and $i=2$ to denote the planet
initially more distant from the star. A mass ratio of $\mu_i \simeq
10^{-3}$ corresponds to $m\simeq1\,M_ {\rm Jup}$ for $M=1\,M_\odot$,
where $M_{\rm Jup}$ is the mass of Jupiter and $M_{\odot}$ is the mass
of the sun.
The initial semimajor axis of the inner planet ($a_{1,\mathrm{init}}$)
was set to unity and the initial semimajor axis of the outer planet
($a_{2,\mathrm{init}}$) was drawn from a uniform distribution ranging
from $0.9 \cdot a_{1,\mathrm{init}} \left( 1+\Delta_c \right) $ to
$a_{1,\mathrm{init}} \left( 1+\Delta_c\right)$, where $1 + \Delta_c$
is the critical semi-major axis ratio above which Hill stability is
guaranteed for initially circular coplanar orbits, and $\Delta_c
\simeq 2.4 \times \left(\mu_1+\mu_2\right)^{1/3}$
(Gladman 1993).  
For small non-zero eccentricities, some of our initial conditions will
result in stable planetary systems.  If we were to calculate branching
ratios for collisions or ejections relative to the total number of
simulations, then our results would depend slightly on our choice of
initial conditions.  Therefore, we present the fractions of systems
that have a certain property, conditioned on there being an ejection.
When presented in this form, our results are insensitive to the number
of systems that remained bound and whether they were chaotic.

The initial eccentricities were distributed uniformly in the range
from 0 to 0.05, and the initial relative inclination in the range from
$0^{\circ}$ to $2^{\circ}$. All remaining angles (longitudes and
phases) were randomly chosen between 0 and $2\pi$.  Throughout this
paper we quote numerical results in units such that
$G=a_{1,\mathrm{init}}=M_{\star}=1$, where $G$ is the gravitational
constant. In these units, the initial orbital period of the inner
planet is $P_1\simeq 2\pi $.

Throughout the integrations, close encounters between any two bodies
were logged, allowing us to use a single set of n-body integrations to
study the outcome of systems with a a wide range of planetary radii.
We consider a range of radii to allows for the uncertainty in both the
physical radius and the effective collision radius allowing for
dissipation in the planets.  When two planets collided, mass and
momentum conservation were assumed to compute the final orbit of the
resulting single planet.

Each run was terminated when one of the following four conditions was
encountered: (i) one of the two planets became unbound (which we
defined as having a radial distance from the star of $2000\,
a_{1,\mathrm{init}}$); (ii) a collision between the two planets
occurred assuming $R_i/a_{1,\mathrm{init}} = R_{\rm
min}/a_{1,\mathrm{init}} = 1\, R_{\rm Jup}/5\,{\rm AU}=
0.95\times10^{-4}$, where $R_{\rm Jup}$ is the radius of Jupiter;
(iii) a close encounter occurred between a planet and the star
(defined by having a planet come within $r_{\rm
min}/a_{1,\mathrm{init}}= 10\,R_\odot/5\,{\rm AU}= 0.01$ of the star);
(iv) the integration time reached $t_{\rm max}= 5\cdot 10^6 - 2 \cdot
10^7$ depending on the masses of the planets.
These four types will be referred to as ``collisions,'' meaning a
collision between the two planets, ``ejections,'' meaning that one
planet was ejected to infinity, ``star grazers,'' meaning that one
planet had a close pericenter passage, and ``two planets.''

\subsection{Results}

\begin{figure}
\epsscale{1.0}
\plotone{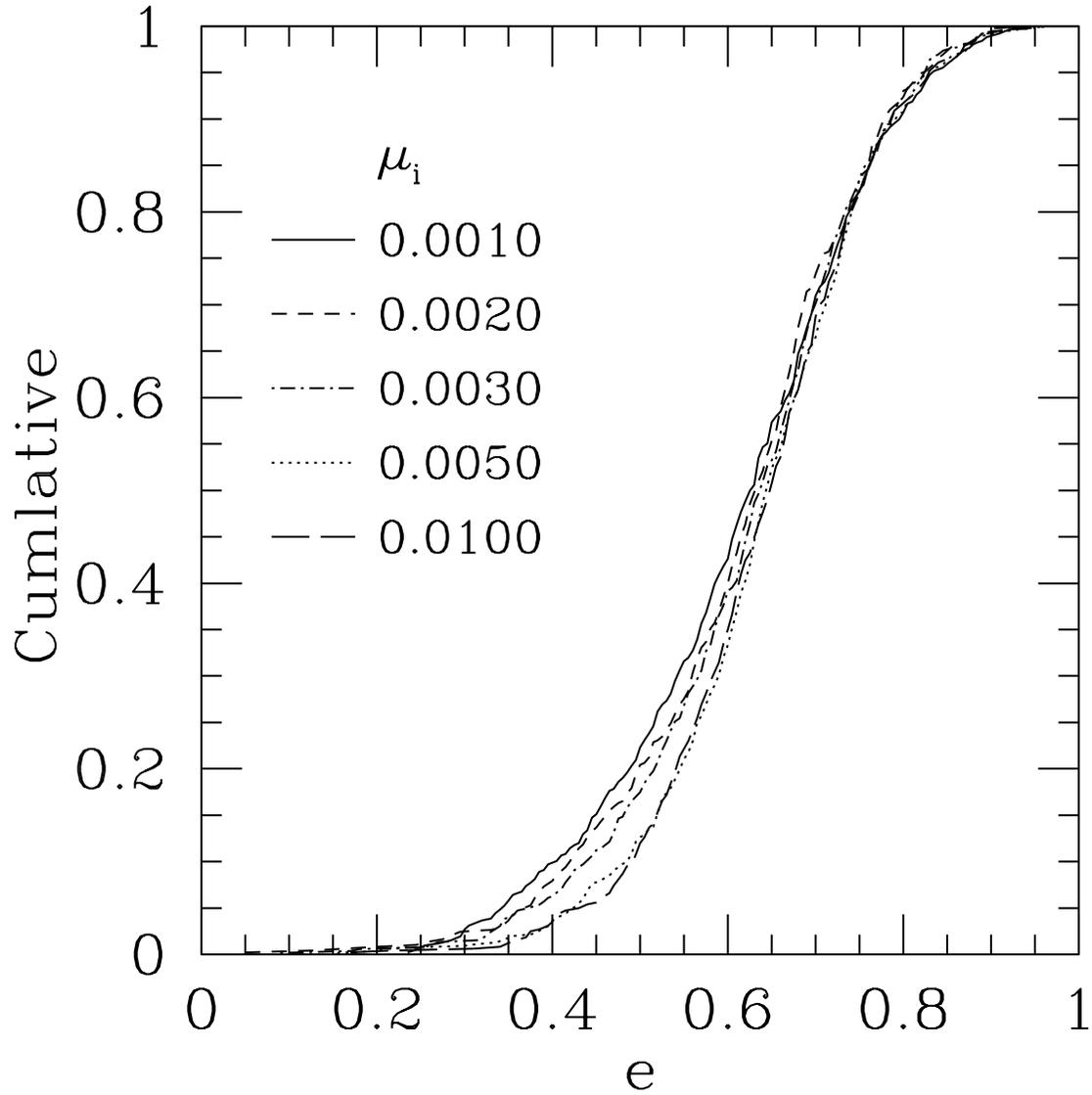} 
\caption{\label{FigEccDistTotalMass}
Cumulative distributions for the final eccentricity of
the remaining planet after the other planet has been ejected.  Each
line style corresponds to an ensemble of simulations with a different
value of $\mu_1=\mu_2$ as specified in the legend.  While the branching
ratios for different outcomes depend on the total planet mass, the orbital
properties for a given type of outcome (i.e., ejection or collision) are
insensitive to the total mass.
}
\end{figure}

We began by conducting an exploratory set of integrations using a wide
variety of planet masses ($10^{-3} \le \mu_i < 10^{-2}$).  The
probabilities for the four outcomes (collisions, ejections, star
grazers, and two planets) depend on the masses of both planets.
However, based on a set of preliminary integrations, we found that the
final orbital properties of the system within one of these outcome types
depend on the ratio of planet masses, but are insensitive to the
total planet mass.  In Fig.\ \ref{FigEccDistTotalMass}, we illustrate
this point by plotting the cumulative eccentricity distribution
following an ejection for six sets of simulations each with a fixed
mass (for both of the two planets).  In light of the insensitivity to
the total mass, we focused our n-body integrations of a series of
seven sets of integrations with a constant total planet mass ratio,
but varying $\beta\equiv~m_{(2)}/(m_1+m_2)$, where we use $m_{(1)}$
and $m_{(2)}$ to denote the mass of the more and less massive planets.
In this set of integrations, we include an expanded range of planet
masses ($10^{-4} \le \mu_i < 10^{-2}$) and choose initial conditions
that include both the more massive planet having a smaller initial
semi-major axis ($m_1=m_{(1)}$) and the less massive planet having a
smaller initial semi-major axis ($m_1=m_{(2)}$).  We choose a somewhat
large total planet mass ratio, $\mu_1+\mu_2 = 6\times
10^{-3}$, so as to accelerate the evolution of the planetary systems
and reduce the computational cost of the simulations.

\subsubsection{Collisions}

Collisions leave a single,
larger planet in orbit around the star.  Near the time of a collision,
the energy in the center-of-mass frame of the two planets is much
smaller than the gravitational binding energy of a giant planet to the
star.  Therefore, we model the collisions as completely inelastic and
assume that the two giant planets simply merge together while
conserving total momentum and mass.  Using this assumption, the final
orbit has a semi-major axis between the two initial semi-major axes, a
small eccentricity, and a small inclination.  In fact, we find that
the final semi-major axis is only slightly less than would be
estimated on the basis of energy conservation,
\begin{equation}
\frac{a_f}{a_{1}} \simeq \left[ \frac{m_1}{m_1+m_2} + \frac{m_2 a_1}{\left(m_1+m_2\right) a_2} \right]^{-1},
\end{equation}
where $a_f$ is the final semi-major axis of the remaining planet.
This compares favorably with the results of our simulations and the
magnitude of the deviations can be approximated (see appendix of FHR).
We find that the fractions of our integrations that result in
collisions decreases for more more extreme planet mass ratios
(assuming constant total planet mass).  While collisions between
planets may affect the masses of extrasolar planets, a single
collision between two massive planets does not cause significant
orbital migration or eccentricity growth if the planets are initially
on low-eccentricity, low-inclination orbits near the Hill stability
limit.  Therefore, we shift our attention to those simulations that
resulted in ejections.

\subsubsection{Ejections}
\label{sec_eject}

Since the escaping
planet typically leaves the system with a very small (positive)
energy (Moorehead \& Adams 2005), energy conservation sets the final semimajor axis of the
remaining planet slightly less than
\begin{equation}
\frac{a_f}{a_{1}} \simeq \left[ \frac{m_1}{m_f} + \frac{a_1 m_2}{a_2 m_f} \right]^{-1}
\end{equation}
Thus, the final semi-major axis of the planet left behind after an
ejection depends on whether the more massive planet initially had the
smaller or larger semi-major axis.  Otherwise, the order of the
planets makes little difference.  Even in simulations with equal mass
planets ($\beta=0.5$), we that the outer planet typically accounts for
$\simeq55\%$ of the ejections.  When $\beta$ is reduced only slightly
to 0.45 or 0.40, then $\simeq65\%$ or 80\% of the ejections are of the
less massive planet, regardless of which planet was initially closer.
For $\beta\le~0.30$, less than 1\% of the ejections leave the less
massive planet bound to the star.
Therefore, we have combined the final eccentricities and inclinations
of integrations with the same mass ratio, but reverse initial ordering
of the planets.  We present the mean and standard deviation of the
final planet's semi-major axis, eccentricity, and inclination for each
set of simulations in Table 1 (based on a total of 6525 numerical
integrations that resulted in one planet being ejected).  Thus, the
ejection of one of two equal-mass planets results in the most
significant reduction in the semi-major axis, but is limited to
$\frac{a_f}{a_{1,\mathrm{init}}} \ge 0.5$.
%


In simulations resulting in an ejection, the remaining planet acquires
a significant eccentricity, but its inclination typically remains
small.  The eccentricity and inclination distributions for the
remaining planet are not sensitive to the sum of the planet masses,
but depend significantly on the mass ratio.  Both the final
eccentricity and inclination are maximized for equal-mass planets.

In Fig.~\ref{fig_evsbeta} we show the cumulative distributions for the
eccentricity after an ejection for different mass ratios.  While any
one mass ratio results in a narrow range of eccentricities, a
distribution of mass ratios would result in a broader distribution of
final eccentricities.  However, there is a maximum eccentricity, which
occurs for equal-mass planets.  Thus, the two-planet scattering model
predicts that eccentricities rarely greatger than $\simeq0.8$, {\em
independent of the distribution of planet masses}.  We will compare
this with the properties of known planets is \S~4.

\subsubsection{Stargrazers} \label{SecStarGrazers}

In a small fraction of our numerical integrations one planet underwent
a close encounter with the central star (i.e., came within
$10^{-2}\times~a_{1,\mathrm{init}}$).  Note that the upper limit for
the eccentricity of the remaining planet of 0.8 applies only after the
other planet has been ejected from the system.  Smaller pericenter
separations can be achieved before a planet is ejected, while the
orbits of both planets contain significant angular momentum.  For our
simulations with $\beta=0.5$, $R_p/a_{1,\mathrm{init}} = 10^{-4}$),
$\simeq~3$\% of all our integrations resulted in a star grazer.  While
the overall fraction of runs that result in stargrazers is sensitive
to the planetary radii, the ratio of the number of integrations that
resulted in stargrazers to the number that resulted in ejections is
not.  Additionally, the ratio of the number of integrations that
resulted in stargrazers to the number that resulted in ejections is
likely to be less sensitive to our choice of initial conditions.  For
the same parameters, we find a ratio of $\simeq~0.06$.  In our
simulations with more extreme mass ratios, we find the total fraction
of runs resulting in a star grazer is $\simeq~12$\% or $\simeq~16$\%
for $\beta=0.3$ or $\beta=0.2$, and the ratio of star grazers to
ejections is $\simeq~0.2$ or $\simeq~0.3$.

We must exercise caution in interpreting the above numbers.  Due to
the limitations of the numerical integrator used, the accuracy of our
integrations for the subsequent evolution of systems resulting in star
grazers cannot be guaranteed (Rauch \& Holman 1999).  Moreover, some
of these planets could be directly accreted onto the star if their
pericenters continue to decrease, or they might be ablated or
destroyed by stellar winds/radiation (Vidal-Madjar et al.\ 2003, 2004;
Murrary-Clay et al.\ 2005), or even ejected from the system following a
strong tidal interaction (Faber, Rasio \& Willems 2005). 
Moreover, the orbital dynamics of these systems might be affected by
additional forces (e.g., tidal forces, interaction with the quadrupole
moment of the star, general relativity; Adams \& Laughlin 2006) that
are not included in our simulations and would depend on the initial
separation and the radius of the star.  For example, Nagasawa et al.\
(2008) find that including tidal forces throughout the simulation can
significantly increase the number of planets circularized in
short-period orbits.  Despite these complications, our simulations can
provide constraints on the frequency of short-period planets formed
via a combination of planet scattering and tidal dissipation.

The fraction of systems producing stargrazers in our simulations is
larger than the fraction of solar-type stars in radial velocity
surveys that have very-hot-Jupiters (1d$\le$P$\le$3d) or hot-Jupiters
(3d$\le$P$\le$5d), but smaller than the fraction of hot-Jupiters among
detected extrasolar planets (Butler et al.\ 2006).  The results of the
OGLE-III transit search allow estimates for the frequency of
hot-Jupiters ($\simeq\left(1^{+1.39}_{-0.59}\right)/310$) and very-hot
Jupiters ($\simeq\left(1^{+1.10}_{-0.54}\right)/690$).  Only the former
rate is statistically consistent with current estimated rates based
on radial velocity surveys ($\simeq0.6$\% for hot-Jupiters; Gould et
al.\ 2006).  While the fraction of solar-type stars with short-period
giant planets is well constrained by existing radial velocity surveys,
the frequency of long-period planets is not yet well constrained.  The
present detections provide a lower limit on their frequency, but this
fraction is expected to increase as radial velocity surveys extend to
longer temporal baselines.  Improvements in measurement precision and
instrument stability will enable the detection of less massive
long-period planets, and is also likely to increase the number of
long-period giant planets.

Given the limitations of our simulations and existing observations, it
is most appropriate to compare: (a) the theoretical ratio of the number
of systems (from our simulations) that resulted in stargrazers to the
number of systems that resulted in ejections to (b) the upper limit for
the current observational ratio of the frequency of
(very-)hot-Jupiters to the frequency of eccentric giant planets.
Restricting our attention to planets with $m\sin i\ge0.1M_J$, we
find 20 planets with orbital periods between 3 and 5 days (not
including those recently discovered by the ``N2K'' project that
focuses on short-period planets) and 80 planets with best-fit
eccentricities greater than 0.2.  Thus, we estimate the upper limit
for the observational ratio to be $\simeq 20/80=0.25\pm0.06$.  Given
the uncertainties in both the observational and theoretical ratios,
this suggests that planet--planet scattering could be responsible for a
significant fraction of hot-Jupiters, if typical planetary systems
form multiple giant planets.  While radial velocity surveys are still
incomplete, at least $\simeq30$\% of stars harboring one giant planet
show evidence for additional for distant giant plants (Wright et al.\
2007).  Thus, our simulations suggest that for giant planets with
initial semimajor axes of a few AU, it is possible to achieve the
extremely close pericenter distances necessary to initiate tidal
circularization around a main sequence star and possibly leading to
the formation of (very-)short-period planets (Rasio \& Ford 1996;
Rasio et al.\ 1996; Faber, Rasio \& Willems 2005). 

The planet scattering plus tidal circularization model for forming
giant planets with very short orbital periods will be tested by future
observations.  Measurement of the Rossiter-McLaughlin effect could
detect a significant relative inclination between the planet's orbital
angular momentum and the stellar rotation axis (Gaudi \& Winn 2007)
that could be induced by planet scattering (Chatterjee et al.\ 2008).
Observations of $11^\circ\pm15^\circ$ for HD 149026 (Wolf et al.\
2007) and $30^\circ\pm21^\circ$ for TrES-1 (Narita et al.\ 2007) are
suggestive and will stimulate additional observations to improve the
measurement precision.  On the other hand, the detection of a
Trojan companion to a short-period giant planet would suggest that
the planet's migration was less violent (e.g., Ford \& Gaudi 2006;
Ford \& Holman 2007).

\subsection{Final Semi-major Axes}
The ejection of one of two giant planets results in an inward
migration of the remaining planet, but this migration is limited by
$\frac{a_f}{a_{1,\mathrm{init}}} \ge 0.5$.  Assuming that giant
planets form (and emerge from the disk) at locations beyond the
ice-line ($\sim 2.7$AU for a solar mass star), two-planet scattering
by itself is unlikely to explain the origins of giant planets with
orbital periods $\ll~1.3$AU.  However, the modest inward migration
caused by two-planet scattering may be responsible for the observed
increase in the frequency of giant planets at orbital periods just
beyond
$\sim$1 year (and hence separations of $\sim$1AU for solar-type stars;
Cumming et al.\ 2008).

A combination of planet scattering and tidal effects may be able to
explain the origins of (very-)hot-Jupiters (\S3.2.3; Nagasawa
et al.\ 2008).  However, the scattering of two giant planets has more
difficulty explaining the presence of giant planets in nearly circular
orbits at intermediate orbital periods $\sim10\,$d--$100\,$d, since
their orbits are small enough to require significant migration, but
large enough that tidal circularization is ineffective.  Additional
physics such as tidal effects and/or disk migration such planets may be necessary
to explain such planets.  For
example, it might be possible to circularize such giant planets at
intermediate distances, if the circularization occurs while the star
is on the pre-main sequence and has a larger radius or while a
sufficiently massive circumstellar disk is still present.

Of course intermediate-period planets may also have formed via some
other mechanism (\S1).  For example, migration prior to instability
could result in giant planets with these intermediate orbital period.
In this case, the initial conditions of our simulations may not be
representative, if two planets started with larger separations and the
migration drove them into a mean motion resonance (e.g., 2:1 or 3:2).
Therefore, more detailed simulations accounting for the planet-disk
interactions are necessary to explore this scenario.  For example,
Adams \& Laughlin (2003) explore pairs of migrating giant planets that
typically become trapped in a 3:1 mean motion resonance, leading to
eccentricity growth and eventually dynamical instability.  Similarly,
Moorhead \& Adams (2005) consider planet scattering in the presence of
type II migration and find that many systems evolve towards the 2:1 or
3:1 mean motion resonance.  Both studies find that planet scattering
plus planet-disk interactions are able to produce giant planets spread
across a broad range of semi-major axes, including separations of less
than $\sim$1AU.  However, these simulations do not produce a
significant increase in the frequency of giant planets just beyond
$\sim1$AU, as inferred from radial velocity observations (Cumming et
al.\ 2008).  In any case, these simulations suggest that such planets
can maintain large eccentricities, despite the presence of
eccentricity damping.  While the detailed assumptions and initial
conditions are different, it is noteworthy that both the idealized
two-planet scattering experiments presented here and the simulations
of Adams \& Laughlin (2003) over-produce highly eccentric planets
(relative to current observations; see \S4.3).

\section{Comparison with Observations}

In this section, we investigate the observed distribution of
eccentricities of extrasolar planets, based on the catalog of Butler
et al.\ (2006), as updated by Johnson et al.\ (2006) and Wright et
al.\ (2007).  For comparing the observed eccentricity distribution to
models, it is useful to exclude some of these planets, when the
eccentricities are only weakly constrained by present observations.
When the time span of radial velocity observations span less than two
orbital periods, there can be significant degeneracies between the
orbital period, eccentricity, and other parameters, and the bootstrap
method of estimating uncertainties in orbital parameters can
significantly underestimate the true uncertainties (Ford 2005).
Therefore, we restrict our attention to planets with an orbital period
less than half the time span of published radial velocity observations.
Similarly, we exclude planets with orbital periods less than 10 days,
since their eccentricities may have been altered due to tidal
dissipation (Rasio et al.\ 1996).  Of the 173 planets discovered by
the radial velocity method, 136 meet both these criteria.  Of these
136, the best-fit eccentricity for 86 planets exceeds 0.2.
The abundance of giant planets with large eccentricities has led
theorists to develop several models for exciting orbital
eccentricities.  Here we consider the implications of the observed
eccentricity distribution for the planet--planet scattering model.
%

In order test this, we compare the eccentricity distribution predicted by the two-planet scattering model to the observed eccentricity distribution.  

\begin{figure}
\epsscale{1.0}
\plotone{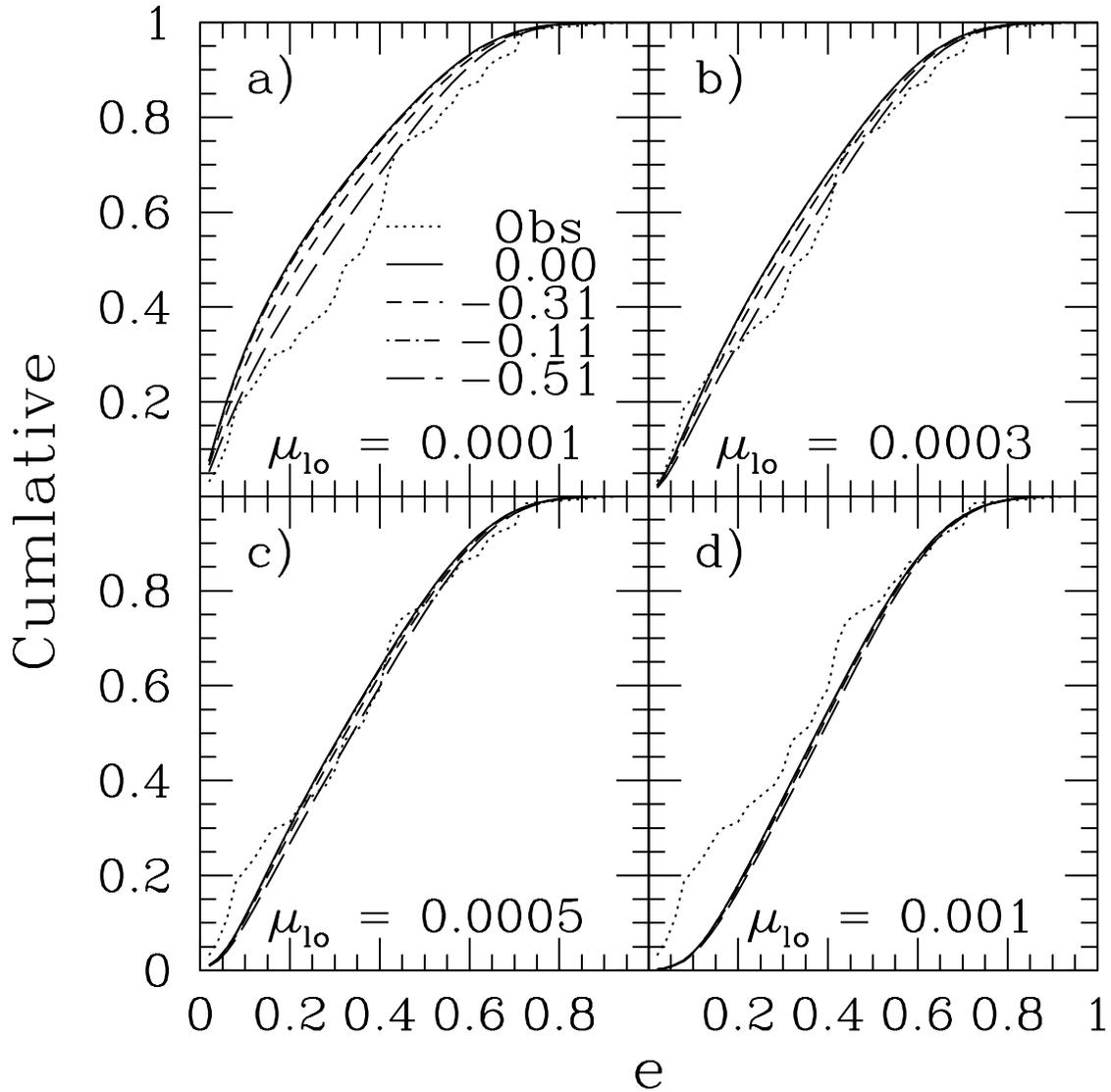} 
\caption{\label{fig_genecc}
Cumulative distributions for the final eccentricity of the remaining
planet after the other planet has been ejected.  Each line style
corresponds to an ensemble of systems with a different power law
index for the mass distribution, as indicated in the legend of panel
a.  The solid line corresponds to an initial distribution uniform in
the log of the planet mass.  The short dashed curves correspond to
best estimate for the planet mass distribution based on the analysis
of Cumming et al.\ 2008.  The short-long dashed and dot-dashed curves
correspond to the limits of the quoted 1-sigma uncertainty in the
power-law index.  The dotted curve in each panel corresponds to the
observed eccentricity distribution (excluding planets with orbital
periods less than 10 days).  The different panels correspond to
different choices for the cutoff at low masses. Each panel assumes a
cutoff of $\mu_{\rm hi}=10^{-2}$, based on findings of radial velocity
surveys. 
}
\end{figure}

We generated tens of thousands of new planetary systems containing two
giant planets, assuming that the planet masses are uncorrelated.  We
assume that each of these systems would result in the less massive
planet being ejected and estimate the final eccentricity of the more
massive planet that remains bound.  To determine the final
eccentricity very efficiently, we use simple analytic fits to the mean
and standard deviation of the eccentricity of the bound planet after
one planet has been ejected, as a function of $\beta$, the planet mass
ratio.  We show the resulting eccentricity distributions in
Fig.~\ref{fig_genecc} for several different planet mass distributions.
Within each panel, the different line styles indicate different
choices of power-law index for the planet mass.  These choices are
based on a recent analysis of the Keck Planet Search estimated the
mass-period distribution of giant planets with periods less than 2000
days.  Cumming et al.\ (2008) found a best-fit mass power-law index of
$0.31\pm0.20$.  Fig.~\ref{fig_genecc} shows that in the two-planet
scattering model, the final eccentricity following an ejection is
relatively insensitive to this power law index (within the range
consistent with observations).  However, we find that the eccentricity
distribution predicted by the two-planet scattering model is more
sensitive to the choice for the lower cutoff of the planet-star mass
ratio ($\mu_{\rm lo}$).  If the typical planetary system forms two giant
planets with masses distributed across two orders of magnitude
($\mu_{\rm lo}=10^{-4}$, $\mu_{\rm hi}=10^{-2}$), then the two-planet
scattering model would over predict low eccentricity planets.  On the
other hand, if the typical planetary system forms two giant planets
with masses distributed across only one orders of magnitude ($\mu_{\rm
lo}=10^{-3}$, $\mu_{\rm hi}=10^{-2}$), then the two-planet scattering
model would over-predict high eccentricity planets.  For intermediate
lower mass ratio cut-offs (e.g., $\mu_{\rm lo}=3\times10^{-3}$ or
$5\times10^{-4}$, the predicted and observed eccentricity
distributions show excellent agreement.
%
%
Our results demonstrate that the exact distribution of eccentricities
predicted will depend on the distribution of planet masses, including
whether the masses of multiple planets that formed around one star are
correlated.  Understanding the distribution of planet masses and
orbits is the subject of much ongoing research.  Therefore, we first
turn our attention to predictions of the planet scattering model that
are insensitive to these uncertainties.  In particular, the planet
scattering model predicts that the planet that remains bound to the
host star following an ejection will typically be the more massive of
the two planets.  As a result, the most extreme final eccentricities
occur as the result of scattering of nearly equal mass planets.  In
the limiting case of equal mass planet scattering, the average final
eccentricity was $\left<e_f\right>=0.624$ and the standard deviation
was $\sigma_{e_f}=0.135$.  Therefore, the planet--planet scattering
model predicts that eccentricities very rarely exceed $\simeq0.8$.  Then,
we investigate what distribution of planet mass ratios would be
necessary to explain the eccentricity distribution derived from
present observations.  Finally, we explore the potential for
correlations between the eccentricity distribution and other
properties to constrain theories for the origin of eccentricities,
including the planet-planet scattering model.

\subsection{Very High Eccentricity Planets}

Of the 86 planets with orbital periods greater than 10 days and
best-fit orbital eccentricities exceeding 0.2, two planets have
eccentricities that are currently estimated to be greater than 0.8, HD
80606b ($e=0.935\pm0.0023$; Naef \etal 2001) and HD 20782b
($e=0.925\pm0.03$; Jones et al.\ 2006).  Such large eccentricities are
unlikely to be the result of planet--planet scattering (at least
in the context of two planets initially on nearly circular orbits, as
explored in \S\ref{sec_eject}).  Thus, we search for alternative
explanations for these two planets with extremely large
eccentricities.  First, we note that the eccentricity determination of
HD 20782b is quite sensitive to a single night's observations.  If the
observations from that night are omitted, then the best-fit
eccentricity would drop to $0.732$, a value consistent with the
planet-planet scattering model.  Clearly, it would be desirable to
obtain several additional radial velocity measurements around the time
of future periastron passages to confirm the very large eccentricity.

Another possibility is that a wide stellar binary companion may have
played a role in exciting such large eccentricities.  Indeed, both of
these planets orbit one member of a known stellar binary (Desidera \&
Barbieri 2007).  For the sake of comparison, we note that only 19 out
of the 86 planets in our sample orbit members of a known binary.  In
principle, secular perturbations due to a wide binary companion on an
orbit with a large inclination relative to the planet's orbit can
induce eccentricity oscillations with amplitudes approaching unity.
However, the timescale for the eccentricity oscillations can be quite
large for wide binaries, in which case other effects (e.g., general
relativity or other planets) may lead to significant precession of the
longitude of periastron and limit the amplitude of the eccentricity
oscillations (Holman et al.\ 1997; Ford et al.\ 2000; Laughlin \& Adams
2006; Takeda et al.\ 2008).  For both HD 80606 and HD 20782, the orbit
of the wide binary companion is unknown, limiting the utility of
n-body integrations for these systems.  Nevertheless, it is still
possible to estimate the secular perturbation timescale based on the
current projected separation of the binary companion.  The current
estimates in both of these systems are quite large (Desidera \&
Barbieri 2007).  This has led to speculation that the ``Kozai effect''
may not be able to explain the large eccentricities for these two
systems.  The binarity may still be significant, e.g., if the two
stars were not born as a binary, but rather the current binary
companion originally orbited another star and was inserted into a wide
orbit around the planetary system via an exchange interaction (a
formation scenario similar to that proposed for the triple system PSR
1620--26; Ford et al.\ 2001; or other putative planets in multiple
star systems; Portegies Zwart \& McMillan 2005; Pfahl \& Muterspaugh
2006; Malmberg et al.\ 2007b).  During such an encounter, the
four-body interactions might have induced a large eccentricity in the
planetary orbit.  Such interactions may have been common for stars
born in clusters or other dense star forming regions (Adams \&
Laughlin 1998; Zakamska \& Tremaine 2004).

We note that four other planets in our sample currently have a large
best-fit eccentricity, but current observational uncertainties imply
that they may or may not be a challenge to the planet-planet
scattering model: HD 45450 ($e=0.793\pm0.053$), HD 2039
($0.715\pm0.046$), HD 222582 ($e=0.725\pm0.012$), and HD 187085
($e=0.75\pm0.1$).  Additionally, some recently discovered planets---
e.g., HD 137510 ($e=0.359\pm0.028$) and HD 10647 ($e=0.16\pm0.22$)---
have modest best-fit eccentricities and formal uncertainty estimates,
but a Bayesian analysis (following Ford et al.\ 2006) of the
observations indicates significant parameter correlations and/or broad
tails that still allow for very large eccentricities.  We encourage
observers to make additional observations of these known planetary
systems, so as to improve the current observational uncertainties.
Many observations spanning multiple periods with a high degree of
long-term stability and good coverage near periastron passage are
especially important for these particularly interesting high
eccentricity systems that may provide insights into additional
mechanisms for eccentricity excitation.  We also encourage observers
to pursue broad planet searches, so as to increase the number of known
planets with very large eccentricities.  Discovering a larger sample
of such planets and follow-up observations help determine the role of
binary companions in forming such systems.

\subsection{Inferred Planet Mass Ratio Distribution}

In \S\ref{sec_numint}, we demonstrated that the planet-planet
scattering model predicts a large distribution of eccentricities and
could account for 84 of 86 planets in the current sample.  Thus, the
planet--planet scattering model might be the dominant mechanism for
exciting the eccentricities of extrasolar planets.  In order to further
explore this possibility, we consider the limiting case in which every
planet's eccentricity is presumed to be due to the planet having
ejected exactly one other planet.  Since the eccentricity of the
remaining planet depends strongly on the mass ratio and the less
massive planet is almost always ejected, we are able to transform the
observed eccentricity distribution into a distribution of the inferred
planet mass ratios (where $\beta_f = m_f/(m_1+m_2)$ is the ratio of
the mass of the putative ejected planet to the sum of masses of that
planet and the remaining planet, assuming the orbits are coplanar).
To perform this inversion, we assume that the final eccentricity is
uniquely determined by the ratio of planet masses, and use the fitting
formula $e_f = 1.44 \beta_f^{1.23}$, based on the median
eccentricities shown in Table~1 and Fig.~\ref{fig_evsbeta}.  Since
this fitting formula is based on the median eccentricities from our
scattering simulations, it is expected that planet--planet scattering
for equal-mass planets would result in some final eccentricities
slightly greater than 0.62, the predicted median eccentricity
evaluating our fitting formula at $\beta_f=0.5$. Therefore our simple
inversion of the fitting formula would result in $\beta_f$ somewhat
greater than 0.5 for some systems, even if the less massive planet had
always been ejected.

\begin{figure}
\epsscale{1.0}
\plotone{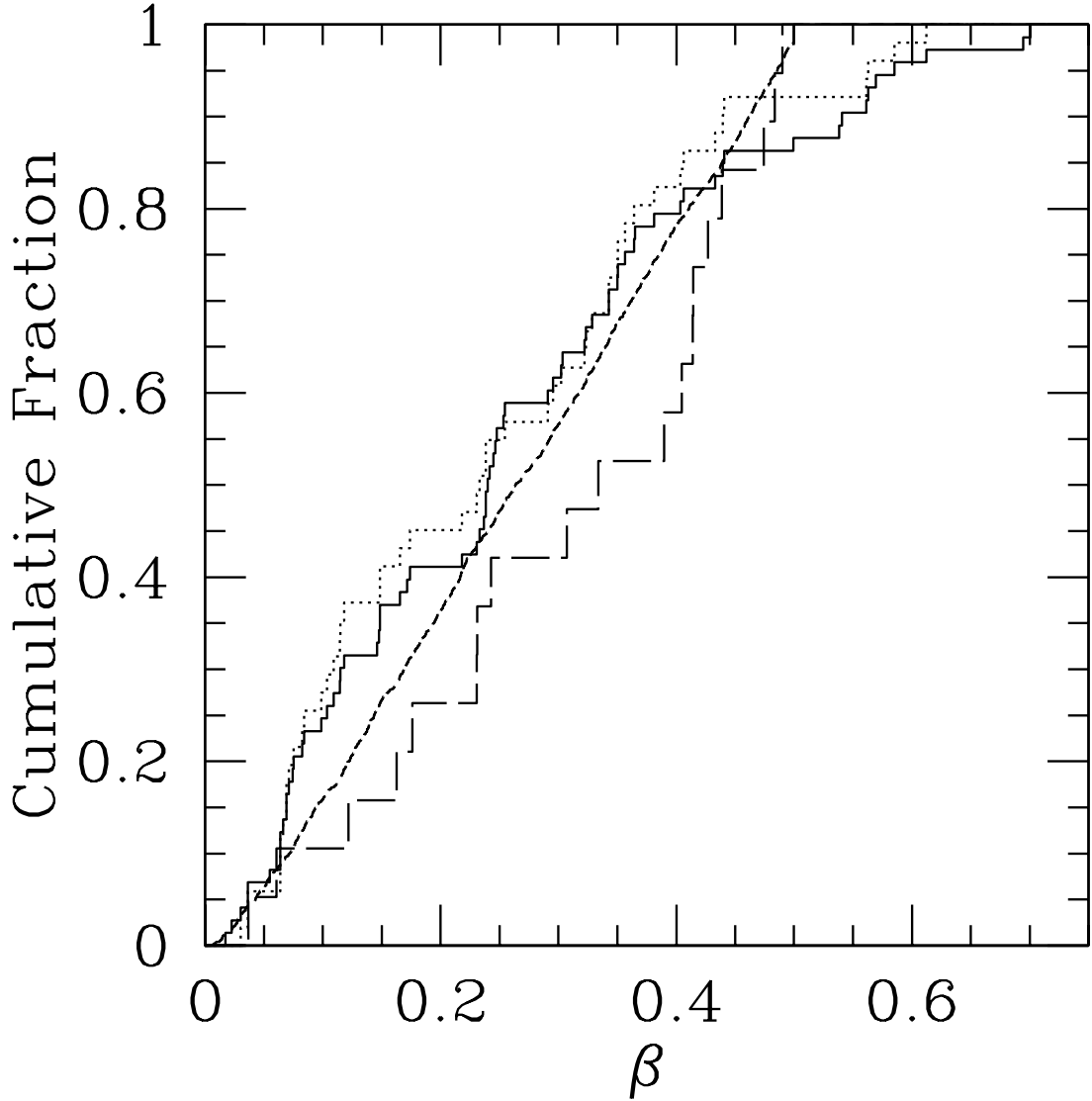} 
\caption{\label{fig_cumbeta}
 Cumulative distributions for $\beta$, the ratio of
the mass of one planet to the sum of the two planet masses.  The solid
curve is determined by assuming that the eccentricity of each
of the known extrasolar planets (with orbital period between ten days
and half the temporal baseline of the published radial velocity
observations) is due to planet-planet scattering.  The dotted line is
similar, but excludes planets in known binary systems.  The long
dashed line is the distribution of $m_{(2)}/(m_{(1)}+m_{(2)})$ for
known multiple planet systems, where $m_{(1)}$ is the most massive
planet known to orbit a star and $m_{(2)}$ is the second most massive
planet orbit known to orbit the same star (assuming coplanar orbits).
The short dashed line is the distribution of $\beta$ derived by
drawing two planet masses (independently and with replacement) from
the catalog of extrasolar planets discovered via radial velocity
planet searches. 
}
\end{figure}

To minimize contamination from either tidally circularized planets or
planets with significant uncertainties in the orbital parameters, we
again base our analysis on the eccentricities of extrasolar planets with
orbital periods greater than 10 days, but less than half the time span
of published radial velocity observations.  We plot the cumulative
distribution of $\beta_f$ for this sample (Fig.\ \ref{fig_cumbeta}, solid
line), as well as a subset of these extrasolar planets where we have
omitted planets in known binary systems (dotted line, 67 planets).  
For comparison, we consider the known multiple-planet systems and plot
the cumulative distribution for $\beta$, the ratio of the second most massive
planet to the sum of the masses of the two most massive (known)
planets in the system (again assuming coplanarity; long dashed line).
The distribution of inferred mass ratios is somewhat more extreme than
the distribution of planet mass ratios for the known multiple planet
systems.  A two-sample Kolmogorov-Smirnov test yields a $p$-value
of 0.19 or 0.14, including or excluding $\beta$'s inferred from the
current eccentricities of planets in known binaries.  One possible
explanation is that planetary systems with a timescale for
instabilities that exceeds the current age of the star could have a
different distribution of $\beta$ than planetary systems that have
already ejected a giant planet.  Another possible explanation is that
the typical history of a planetary system that currently contains a
single giant planet might differ from the history of the typical
planetary system that now has multiple giant planets.  Despite these
possibilities, we caution that any differences in the observed and
inferred distributions of $\beta$ and $\beta_f$ may be due to observational
selection effects.  If a system has a small $\beta$, then one planet
will typically have a much smaller velocity semi-amplitude and be more
likely to have evaded detection.

As another point of reference, we show the cumulative distribution for
$\beta$ that would result from randomly choosing pairs of planets from
our sample (which excludes planets with orbital periods less than 10
days or longer than half the time span of published radial velocities,
but includes planets in known binaries; short-dashed curve).  We
observe that this distribution is quite similar to the distribution of
inferred $\beta_f$'s, differing only for $\beta\ge0.45$.  A two-sample
Kolmogorov-Smirnov test yields a $p$-value of 0.21 or 0.06, depending
on whether we include or exclude planets in known binary systems.  In
principle the difference might be due to nature rarely forming very
nearly equal mass planets.  However, we regard it as more likely that
this difference is due to our assumptions that the less massive planet
is always ejected and that the final eccentricity of the remaining
planet is exactly determined by $\beta$ (i.e., we ignore the
dispersion of $e_f$ observed in our scattering experiments).
Clearly, this comparison is affected by the observational selection
effects that favor detecting massive planets.  Nevertheless, on the
whole, this suggests that the planet scattering model can easily
reproduce the observed eccentricity distribution (for planets with
orbital periods greater than $10\,$d) by assuming a plausible mass
distribution and no strong correlation between the masses of planets
in multiple planet systems.


\subsection{Observed Eccentricity Distribution}
\label{sec_obse}

Next, we analyze the observed distribution of extrasolar planet
eccentricities, without assuming that large eccentricities are due to
planet scattering.  As before, we restrict our attention to the
extrasolar planets with orbital periods between 10 days and half the
time span of published observations.  We have performed a Bayesian
analysis for each of the planets in the catalog of Butler et al.\
(2006), using the radial velocity data sets published by California
and Carnegie Planet Search team.  A detailed analysis will be
presented separately, and here we only summarize our method.  We
assume the published model type (i.e., the number of planets and
whether there is a long term trend) and apply the Markov chain Monte
Carlo algorithm described in Ford (2005, 2006).  Previous work has
shown that the bootstrap-style estimates of parameter uncertainties
employed by Butler et al.\ (2006) can differ significantly from
uncertainty estimates based on the posterior probability distribution
for model parameters (Ford 2005; Gregory 2005, 2006).  Such
differences are common for planets with eccentricities that are very
near 0, planets with orbital periods comparable to the time span of
observations, and planets with few and/or low signal-to-noise
observations. Here, we focus out attention on the marginal posterior
probability distributions for the orbital eccentricities.  By
restricting our attention to planets with orbital period greater than
10 days and less than half the timespan of observations, we obtain a
sample for which the eccentricities are typically well-constrained by
the observations and the two methods typically give qualitatively
similar uncertainty estimates.  We consider individually the multiple
planet systems containing one or more planets with intermediate
orbital periods and one planet with an orbital period likely longer
than half the time span of observations.  We determined that the
uncertainty in the orbital parameters of the outer planet in the
systems Hip 14810, HD 37124, and HD 190360 may significantly affect
the orbital parameters of the other planets.  Therefore, we dropped
all planets in these systems from our sample.

\begin{figure}
\epsscale{0.7}
\plotone{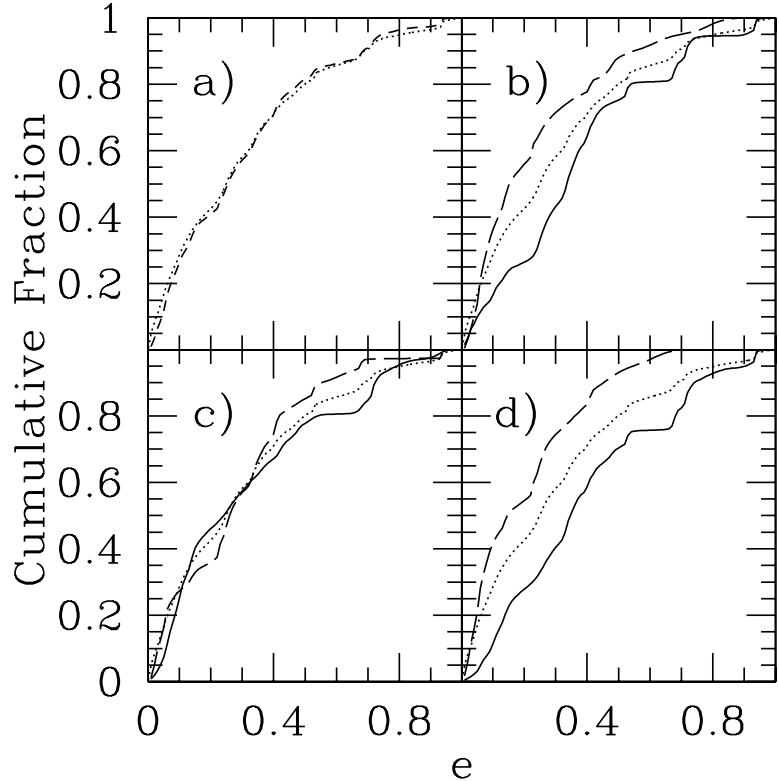} 
\caption{\label{fig_cume_wbin}
Cumularive distributions for eccentricities of
known extrasolar planets.  Each cumulative distribution is based on
the currently known extrasolar planets discovered by radial velocity
surveys with orbital period between ten days and half the temporal
baseline of the published radial velocity observations.  A few planets
have been omitted due to significant uncertainty in the eccentricities
(\S\ref{sec_obse}).  The short-dashed curve in panel a is based on the
published best-fit eccentricities and uncertainties.  The dotted curves
(repeated in each panel) are based on a Bayesian analysis of the
published radial velocity observations for each planetary system
(Butler et al.\ 2006, Johnson et al.\ 2006, Wright et al.\ 2006).  In
panels b-d, we repeat this analysis for two subsets of this sample.
In panel b, the solid (dashed) curve is for planets with $m \sin i$
greater (less) than $1.57M_J$.  In panel c, the solid (dashed) curve is
for planets with an orbital period greater (less) than 350d.  In panel
d, the solid (dashed) curve is for planets with $\theta^2 \sin i$
greater (less) than $1.69$, where $\theta$ is the ratio of the escape
velocity from the planet (evaluated at its surface) to the escape
velocity from the sun (evaluated at the planet's apastron).  The
threshold values of $m\sin i$, period, and $\theta$ were chosen to
result in sample sizes as nearly equal as possible.  
}
\end{figure}

\begin{figure}
\epsscale{1.0}
\plotone{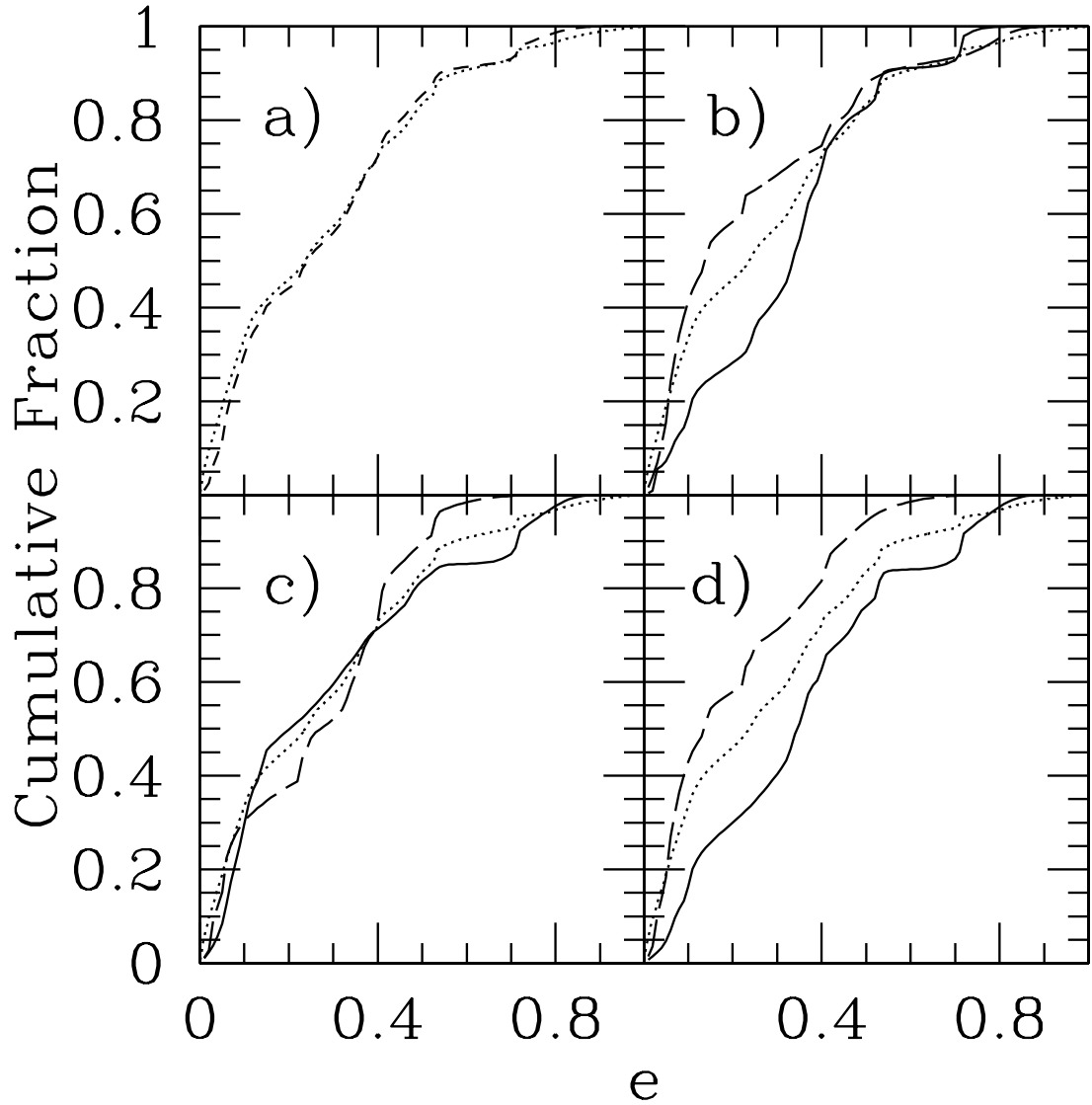} 
\caption{\label{fig_cume_wobin}
Cumularive distributions for eccentricities of
known extrasolar planets not in a known binary system.  Same as Fig.\
4, except only for planets that are not part of a known binary star
system. 
}
\end{figure}

To summarize the available information about the eccentricity
distribution of extrasolar planets, we have averaged the marginal
cumulative posterior eccentricity distribution for each planet in our
sample (Figs.\ \ref{fig_cume_wbin} \& \ref{fig_cume_wobin}, dotted curve,
all panels) for the sample including and excluding planets orbiting
known binary stars.

It is important to note that this method does not provide a Bayesian estimate
of the eccentricity distribution of the population.  Instead, these summary
distributions can be intuitively thought of as a generalization of the
classical histogram that accounts for the uncertainties in the
individual eccentricities in a Bayesian way (allowing for non-Gaussian
posterior distributions).  However, like classical histograms, our
summary distributions can be affected by biases (e.g., the terminal
age bias for dating field stars with stellar models; Pont \& Eyer
2004; Takeda et al.\ 2006).  While we have attempted to minimize the
potential influence of any systematic biases (by selecting a subset of
extrasolar planets for which the eccentricities were well constrained
by observational data), our summary distributions are still influenced
by the shape of individual posterior distributions.  Performing a
proper Bayesian population analysis would require more sophisticated
and much more computationally demanding calculations (e.g., Ford \& Rasio 2006).
Nevertheless, we believe that these distributions can serve as a
valuable summary of the available information about the eccentricity
distribution of extrasolar planets.

For the sake of comparison, we present similar summary 
information for the observed eccentricity distribution based on the
orbital determinations of Butler et al.\ (2006).  For this purpose, we
approximate each planet's marginal cumulative eccentricity
distribution as
\begin{equation}
p(e) = \frac{\erf\left(\left(e-e_{\rm bf}\right)/\sigma_e\right)-\erf\left(-e_{\rm bf}/\sigma_e\right)}{\erf\left(\left(1.-e_{\rm bf}\right)/\sigma_e\right)-\erf\left(-e_{\rm bf}/\sigma_e\right)},
\end{equation}
where $e_{\rm bf}$ is the best fit eccentricity and $\sigma_e$ is the
uncertainty in the eccentricity, both taken from Butler et al.\
(2006).  The results are presented in Figures \ref{fig_cume_wbin}a and 
\ref{fig_cume_wobin}b, dashed curve) for the sample including and
excluding planets orbiting known binary stars.  The strong similarity
of these two distributions demonstrates that this distribution is
well determined and that this can be used as a robust summary of the
observed planet eccentricities.

\subsubsection{Does the Eccentricity Distribution Vary with Planet Mass?}

Next, we investigate whether the eccentricity distribution is
correlated with other planet properties.  Such differences have the
potential to provide insights into the processes of planet formation.
For example, Black (1997) and Stepinsky \& Black (2000) noted
similarities in the period-eccentricity distributions of extrasolar
planets and binary stars, and suggested that both sets of objects may in fact be one
extended population.  More recent work identified differences in the
two distributions and favors the hypothesis that these two populations
have different formation mechanisms (Halbwachet, Mayor \& Udry 20005).

Most recently, Ribas \& Miralda-Escude (2006) noted a potential
correlation between a planet's mass and its orbital eccentricity.
They propose that this could be due to two different formation
mechanisms, e.g., core accretion followed by gas accretion dominating
the formation of planets with $m\sin i\le 4M_J$ and direct collapse of
gas from the protoplanetary nebulae dominating the formation of
planets with $m\sin i\le 4M_J$.  Ribas \& Miralda-Escude (2006)
divided their sample according to $m\sin i$ being greater than or less
than $4M_J$, and test the null hypothesis that the two samples came
from the same distribution using the two sample Kolmogorov-Smirnov
test.  Their choice of $4M_J$ was based on the claimed separation of
exoplanets into two populations based an apparent planet mass-stellar
metallicity correlation.  Since this distinction is only marginal, we
worry that their choice of $4M_J$, we caution against
over-interpretation of their $p$-values that can occur due to {\em a
posteriori} statistical analyses.

We explore this hypothesis by comparing the eccentricity distributions
of various subsets of the extrasolar planet population (presented in aggregate 
as the dotted curve in each panel of Fig.\
\ref{fig_cume_wbin} \& \ref{fig_cume_wobin}).  First, we divide the planet
sample according to the best-fit $m\sin i$ (Fig.\ \ref{fig_cume_wbin}b).
In order to avoid complications associated with {\em a posteriori}
statistics, we choose to perform a single statistical test, dividing
our sample into two nearly equal sized subsamples (they differ in sample size by one):
$m\sin i\le1.57M_J$ (dashed curve) and $m\sin i>1.57M_J$ (solid
curve).  Similar eccentricity distributions that do not include any
planets in known binary systems are presented in Fig.\
\ref{fig_cume_wobin}b.  The same choices of mass ranges results in
subsample sizes that differ by at most three.  A two-sample
Kolmogorov-Smirnov test results in $p$-values of 0.024 and 0.093 for
the samples including and excluding planets in binaries.  (If we had
instead divided our sample at $m\sin i=4M_J$, we would have obtained
$p$-values of 0.004 and 0.023.)  By analyzing planets in systems with
no known binary companion, we minimize the potential for eccentricity
excitation by a stellar binary.  Since the K-S test only suggests
a marginally significant difference between the high- and low-mass 
planets orbiting stars with no known binary companion, we
conclude that a larger sample of extrasolar planets is necessary to
test this hypothesis.  

The sign of any putative correlation between planet mass and
eccentricity is also notable.  It is the more massive planets that are claimed to be
more eccentric.  Many models of eccentricity excitation would predict
that it is easier to increase the eccentricity of lower-mass planets,
since a larger torque is required to excite the eccentricity of a more
massive planet.  One possible explanation is that most planetary
systems produce eccentric giant planets, but the amount of subsequent
eccentricity damping varies from one system to another.  If the late
stage eccentricity damping is determined by the mass of the
planetesimal disk relative to the planet mass, then this model could
explain the larger eccentricities of more massive planets.  Further,
the large dispersion in the time until the
onset of dynamical instability would result in a large dispersion in
the amount of eccentricity damping after the most recent strong
planet scattering event, and thus provide a natural mechanism
for explaining both the small eccentricities of the planets in the
solar system and the large eccentricities of extrasolar giant planets.
Furthermore, this scenario would predict that more massive planets would
tend to have larger eccentricities.

\subsubsection{Does the Eccentricity Distribution Vary with Orbital Period?}

Next, we present a similar analysis, but dividing our planet sample
according to the best-fit orbital period (Figs.\ \ref{fig_cume_wbin}c
\& \ref{fig_cume_wobin}c), rather than $m\sin i$.  A difference in
these distributions might be expected if eccentricity excitation is
strongly correlated with planet migration (Artymowicz 1992; Papaloizou
\& Larwood 2000; Papaloizou \& Terquem 2001; Goldreich \& Sari 2003;
Ogilvie \& Lubow 2003; Cresswell \& Nelson 2007).  If we assume that
giant planets form at large distances and migrate inwards, then
planets that are currently have smaller semi-major axes would be
expected to have experience more migration.  To test this hypothesis,
we divide our sample into two subsets: $P>350$days (solid curve) and
$P<350$days (dashed curve).  Again, the boundary between the two
subsamples is chosen so that size of the two sub samples are equal or
differ by only one when we include binaries and three when we exclude
binaries.  Clearly, the distributions are quite similar.  Formally, a
two sample K-S test results in $p$-values of 0.87 and 0.96 for the
samples that include and exclude planets in binary systems.  Thus, we
conclude that the current planet sample contains no significant
differences in the eccentricity distributions of planets with orbital
periods of 10d$\le~P\le330$d and those with 330d$\le~P\le~T_{obs}/2$,
and we find no observational support for eccentricity excitation via
migration.

It is natural to ask if the large torques presumed responsible for
orbital migration could also be responsible for exciting orbital
eccentricities.  
While the dissipative nature of a gaseous disk naturally leads to
eccentricity damping (Artymowicz 1993), a few researchers have
suggested that excitation may also be possible.  Artymowicz (1992)
found that a sufficiently massive giant planet ($\ge$10 $M_{\mathrm{
Jup}}$) can open a wide gap, leading to torques which excite
eccentricities.  More recently, Goldreich \& Sari (2003) have
suggested that a gas disk could excite eccentricities even for less
massive planets via a finite amplitude instability.  This claim is
controversial, as 3-d numerical simulations have not been able to
reproduce this behavior (e.g., Papalouizou et al.\ 2001; Ogilvie \&
Lubow 2003; Moorhead \& Adams 2008).  Given the large dynamic ranges involved and the 
complexity of the simulations, one might question the accuracy of
current simulations.  For example, 3-d simulations have
suggested that the gaps induced by giant planets might not be as well
cleared as assumed in many 2-d disk models (Bate et al.\ 2003;
D'Angelo et al.\ 2003).  We believe that further theoretical and
numerical work is needed to better understand planet--disk
interactions.  In the meantime, we look to the observations for
guidance on the question of eccentricity damping or excitation.

In the GJ876 system, the observed eccentricities are not consistent
with eccentricity excitation via interactions with the disk.  The
current observed eccentricities could be readily explained if
interactions with a gas disk led to strong eccentricity damping $K =
\dot{e} a / e \dot{a} \gg1$ (Lee \& Peale 2002; Kley et al.\ 2005).
This is in sharp contrast to current hydrodynamic simulations of
migration that suggest $K\simeq1$ and theories that predict $K<0$
(e.g., Goldreich \& Sari 2003; Ogilvie \& Lubow 2003).  While other
resonant planetary systems are not yet as well constrained or studied as GJ
876, the moderate eccentricities of other extrasolar planetary systems
near the 2:1 mean motion resonance suggest that GJ 876 is not unique.

The $\upsilon$ And system also provides a constraint on eccentricity
excitation during migration.  If the outer two planets migrated to
their current locations (0.8 and 2.5AU), then they must have been in
nearly circular orbits at the time of the impulsive perturbation in
order for the middle planet's eccentricity to periodically return to
nearly zero (Ford et al.\ 2005).  While this does not demonstrate a
need for rapid eccentricity damping as in GJ 876, it is inconsistent
with models that predict significant eccentricity excitation.  Since
dynamical analyses severely limit the possibility of eccentricity
excitation in both the GJ 876 and $\upsilon$ And systems, we conclude
that orbital migration does not typically excite eccentricities, at
least for a planet-star mass ratio less than $\sim0.003-0.006$ (those
of the most massive planet in $\upsilon$ And and GJ 876).

\subsubsection{Does the Eccentricity Distribution Vary with the Ability of the Planet to Eject Lower-Mass Objects?}

The next investigation is motivated by theoretical models for
eccentricity excitation via planet-planet scattering and the dynamical
relaxation of packed planetary systems (see \S\ref{sec_intro} and
references therein).  Migration due to scattering of a planetesimal
disk might also result in eccentricity growth for very massive giant
planets, $m_p/M\ge0.01$; Murray et al.\ 1998).  In each of these
models, close encounters can result in either two bodies colliding
(resulting in a more massive planet, but not significant eccentricity
growth) or one body being ejected (resulting in eccentricity growth
for the remaining planet).  The frequency of these two outcomes
depends on
\begin{eqnarray}
\theta^2 & \equiv & \left(\frac{G m}{R_p}\right) \left(\frac{r}{G M_{\star}}\right) \\
 & = & 10 \left(\frac{m}{M_J}\right) \left(\frac{M_\odot}{M_{\star}}\right) \left(\frac{R_J}{R_p}\right) \left(\frac{r}{5 AU}\right),
\end{eqnarray}
where $R_p$ is the radius of the planet (or the effective radius for collision), $r$ is the distance separating the star from the planet at the
time of the close encounter.  Since we do not know the exact distance
for $r$, we set it equal to the current apastron distance of the
observed planet, $a(1+e)$.  When $\theta\gg1$, the planet is able to
efficiently eject bodies, but when $\theta<1$, collisions will be much
more frequent.

We investigate whether the eccentricity distribution is correlated
with $\theta$, by dividing the planet sample according to the ratio of
the escape velocity from the planet (evaluated at the surface of the
planet) to the escape velocity from the star (evaluated at the
apastron distance of the planet) in Figures\ \ref{fig_cume_wbin}d and
\ref{fig_cume_wobin}d.  Since radial velocity observations measure
$m\sin i$, we compare two subsamples with $\theta^2 \sin i\ge 1.69$
(solid line) and $\theta^2\sin i<1.69$ (dashed line).  Again, the
samples are divided such that the equal numbers of planets in each
subsample differs by only one.  Since planets with large $\theta$
eject other bodies more efficiently, the planet scattering model
predicts that massive planets are more likely to acquire large
eccentricities.  This expectation is consistent with the sign of any
putative correlation between the planet eccentricity and $\theta$ seen
in Figures\ \ref{fig_cume_wbin}d and \ref{fig_cume_wobin}d.  A two-sample
K-S test results in $p$-values of 0.020 and 0.100 for samples that
include or exclude planets in known binary systems.  Thus, we find
that the putative dependence of the eccentricity distribution on
$\theta$ is essentially just as statistically significant as the
suggested dependence on planet mass (Butler et al.\ 2006; Ribas \&
Miralda-Escude 2006).  Nevertheless, we caution that both putative
correlations are at most marginally significant at this time.  More
importantly, the similar statistical significance of two putative
correlations of the eccentricity distribution with the planet mass or
$\theta$ demonstrate that even when one model correctly predicts a
correlation, other models may make similar predictions.  Thus, it is
important that theorists explore the implications of a broad range of
theoretical models and observers provide observations that can test
each of these predictions.

We conclude that a larger sample of extrasolar planets would be
valuable for testing hypotheses about the origins of large
eccentricities.  The current sample of $\simeq~200$ extrasolar
planets, has allowed several particularly interesting systems to
provide valuable constraints on planet formation and eccentricity
excitation.  Additionally, the current sample is suitable for
identifying (or refuting) strong correlations, such as the planet
frequency-stellar metallicity correlation (Fisher \& Valenti 2005).
However, many other statistical analyses of the extrasolar planet
population will require the discovery of many more extasolar planets.
While transit searches have the potential to discover many additional
extrasolar giant planets in the coming years, ground-based transit
surveys will be strongly biased towards short-period planets for which
tidal dissipation is likely to have circularized their orbits (Rasio
et al.\ 1996).  These will certainly be quite valuable for testing
theories of planet migration (e.g., Ford \& Rasio 2006) and planetary
structure (e.g., Bodenheimber, Laughlin \& Lin 2003).  However,
statistical investigations of eccentricity excitation mechanisms will
require discoveries many planets with intermediate to long-period
orbits.  Thus, we encourage observers to apply other planet search
techniques to large samples (e.g., ``N2K'' project; Fischer et
al. 2004).  Fortunately, the CoRoT and Kepler space missions will be
able to detect transiting planets with larger separations where tidal
circularization is not significant.  It will be particularly
interesting to study the eccentricities of any rocky transiting
planets found by CoRoT and Kepler.  While it will be extremely
challenging to measure the eccentricity of distant rocky planets via
the radial velocity method, it will be possible to characterize the
eccentricity distribution of such planets (in a statistical sense)
even without radial velocity observations (Ford et al.\ 2008).

\subsection{Could Binaries and Planet Scattering Explain all Large Eccentricities?}

\begin{figure}
\epsscale{1.0}
\plotone{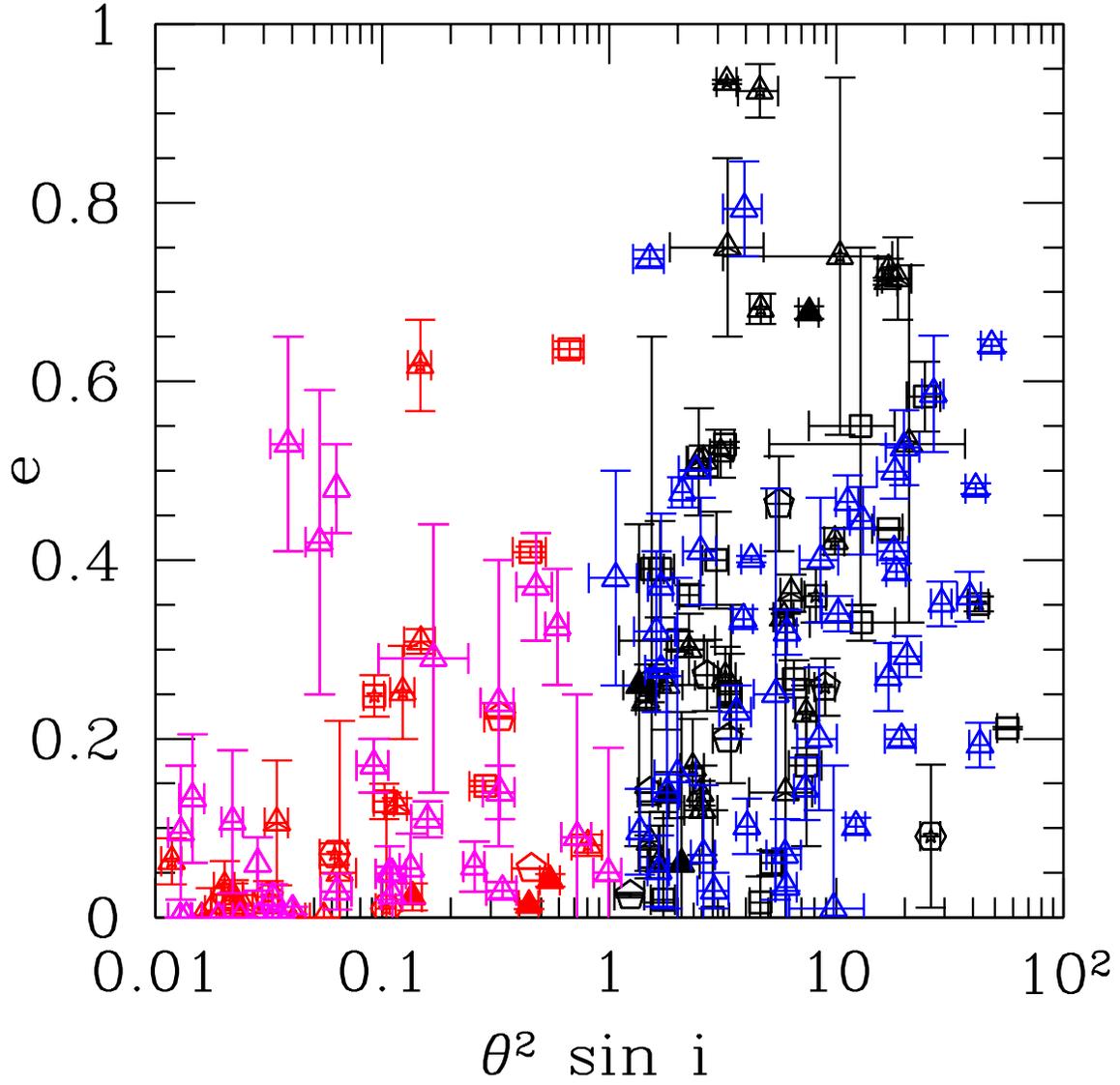} 
\caption{\label{fig_thetavse}
Ratio of the escape velocity from the planet to the
escape velocity from the sun ($\theta^2$) versus eccentricity.  Each
point corresponds to the best-fit solution and uncertainty published in
Butler et al.\ (2006), as updated by Johnson et al.\ (2006) \& Wright
et al.\ (2006).  The planet radius is estimated using the models of
Bodenheimer et al.\ (2003). The horizontal error bar assumes that the
planet radius is known precisely.  The point styles and colors are the
same as in Fig.\ 1.  
}
\end{figure}

Both secular perturbations from a binary companion and planet-planet
scattering appear very likely to play a significant role in exciting
the eccentricities of extrasolar planets, but it is not clear if
additional mechanisms commonly excite eccentricities.  Therefore, we
searched through the catalog of known extrasolar planets to identify
those with large eccentricities that appear unlikely to be due to
either secular perturbations from a wide binary companion or planet
scattering.  We summarize this information in Fig.\
\ref{fig_thetavse}.  Of the 173 radial velocity planets, 89 have
best-fit eccentricities greater than 0.2.  Of those 89, 23 have a
known binary companion and 74 have $\theta \sin i>1$.  This leaves 11
eccentric planets for which the eccentricity cannot be explained by
secular perturbations by a known wide binary companion or by
planet--planet scattering involving that planet.  In three of these
cases, the eccentricity may still have been excited by the
perturbation from another giant planet in the system.  We discuss each
of these briefly.  HD 74156 hosts a second planet that is both
eccentric and capable of ejecting other objects ($\theta^2 \sin
i\simeq 25$). Radial velocities of HD 118203 reveal a long-term trend
that is likely to due to a second more distant giant companion.  While
the orbit is highly uncertain, the slope and time-span of observations
suggest that the putative second planet is most likely to be more
massive than HD 118203b and have a semi-major axis of at least
$1.7\,$AU.  If true, then the putative distant giant planet would be
able to eject other objects and excite an eccentricity in HD 118203b.
Alternatively, the best-fit eccentricity may be an artifact due to the
radial velocity perturbations of one or more additional planets.
Observations also show a long-term trend in HD 49674.  While the
magnitude is smaller, the longer time span of radial velocity
observations imply that the putative planet most likely has an orbital
period beyond $5\,$AU, so it too is likely able to eject other
objects.  GJ 876 contains three planets and the outer two are
participating in a 2:1 mean motion resonances.  Detailed modeling of
this system suggests that the eccentricities of both GJ 876b and GJ
876c are likely due to eccentricity excitation that occurred due to
convergent migration and resonance capture (Lee \& Peale 2002; Kley et
al.\ 2004).  Technically, GJ 876c is massive enough to eject other
bodies ($\theta^2 \sin i\simeq 1.2$), so a hybrid scenario of planet
scattering and resonant capture is possible (e.g., Sandor \& Kley
2006).

This leaves 7 out of 89 eccentric planets that cannot be explained by
secular perturbations from a known wide binary or by planet--planet
scattering by any planet which is currently supported with radial
velocity observations (HD 33283b, HD 108147b, HD 117618b, HD 208487b,
HD 216770b, Hip 14810c), unless $\sin i\ll~1$.  Of course, these
systems may have an undetected binary companion.  For example, there
is preliminary evidence for a binary companion to HD 52265 (Chauvin et
al.\ 2006), but follow-up observations are needed to confirm this.
Similarly, there may be additional undetected planets.  We note that
over half of these systems were discovered only in the last two years, most
have a relatively modest number of radial velocity observations, and
over half of these have a relatively modest signal-to-noise ratio
(velocity semi-amplitude over the effective single measurement
precision).

Thus, there is a very real possibility that additional radial velocity
observations may result in revisions to the measured eccentricity
(e.g., GJ 436b), the detection of a long-term trend most likely due to a
distant giant planet, and/or the detection of additional planets.  The
uncertainties in the orbital elements can be particularly problematic
in cases where there is an undetected planet is near a 2:1 or 3:1 mean
motion resonance.  Previous experience has taught that the
perturbations from yet undetected planets can lead to significant
overestimates of the eccentricity.  We encourage 
additional observations of these systems, which may prove particularly
interesting for testing theories of eccentricity excitation.  We note
that HD 108147, Hip 14810, HD 33283, HD 52265, and HD 216770 are
particularly favorable, since they have velocity amplitudes and
eccentricities such that one eccentric planet could be differentiated
from two planets in a mean motion resonance.  We will present a more
detailed discussion of resonant systems in a future paper.  If future
observations were to confirm the sizable eccentricity, the lack of
other massive companions (both giant planet and stellar companions),
then these systems would provide evidence for at least one additional
eccentricity excitation mechanism in addition to planet-planet
scattering, secular perturbations from binaries, and resonant capture.

\section{Conclusions}
\label{sec_concl}
A planetary system with two or more giant planets may become
dynamically unstable, leading to a collision or the ejection of one of
the planets from the system.  Early simulations for equal-mass planets
revealed discrepancies between the results of numerical simulations
and the observed orbital elements of extrasolar planets.  However, our
new simulations for two planets with {\em unequal masses\/} show a
reduced frequency of collisions as compared to scattering between
equal-mass planets and suggest that the two-planet scattering model
can reproduce the observed eccentricities with a plausible
distribution of planet mass ratios.

Additionally, the two-planet scattering model predicts a maximum
eccentricity of $\sim 0.8$, which is independent of the distribution
of planet mass ratios.  This predicted eccentricity limit compares
favorably with current observations and will be tested by future
planet discoveries.  The current sample of extrasolar planets
provides hints of a correlation between the eccentricity distribution
and other properties.  We show that the putative correlation between
eccentricity distribution and the ratio of the escape velocity from
the planet to the escape velocity from the star is comparable in
statistical significance to the putative correlation between the
eccentricity distribution and planet mass.  Additionally, both
correlations remain marginally significant, when we exclude planets in
known binary systems.  We have identified a few particularly
interesting planets that are unlikely to be explained by the
two-planet scattering model, since $\theta^2 \sin i<1$.  We encourage
additional observations of these systems to determine if these are
isolated planets or if there may be other planets in the system that
could excite large eccentricities and/or cause the current
eccentricity to be overestimated.

The combination of planet--planet scattering and tidal circularization
may be able to explain the existence of giant planets in very short
period orbits.  However, the presence of giant planets at slightly
larger orbital periods (small enough to require significant migration,
but large enough that tidal circularization is ineffective) is more
difficult to explain.  Finally, the planet--planet scattering model
predicts a significant number of extremely loosely bound and 
free floating giant planets, which may also be observable (Lucas \& Roche 2000;
Zapatero Osorio \etal 2002; Veras et al.\ in prep).

A complete theory of planet formation must explain both the eccentric
orbits prevalent among extrasolar planets and the nearly circular
orbits in the Solar System.  Despite significant uncertainties about
giant planet formation, the leading models for the formation and early
dynamical evolution of the Solar System's giant planets agree that the
giant planets in the Solar System went through a phase of large
eccentricities (Levison et al.\ 1998; Thommes et al.\ 1999, 2002;
Tsiganis et al.\ 2005; Ford \& Chaing 2007).  If Uranus and Neptune
formed closer to the Sun, then close encounters are necessary to
scatter them outwards to their current orbital distances.  During this
phase, their eccentricities can exceed $\simeq0.5$ (Tsiganis et al.\
2005).  Alternatively, if Uranus and Neptune were able to form near
their current locations, then oligarch growth predicts that several
other ice giants should have formed contemporaneously in the region
between Uranus and Neptune (Goldreich, Lithwick \& Sari 2004).  The
scattering necessary to to remove these extra ice giants would have
excited sizable eccentricities in Uranus and Neptune (Ford \& Chaing
2007; Levison \& Morbidelli 2007).  Finally, the gravitational
instability model predicts that most giant planets form with
significant eccentricities (Boss 1995).  Therefore, it seems most
likely that even the giant planets in our Solar System were once
eccentric.


Perhaps the question, ``What mechanism excites the eccentricity of
extrasolar planets?'' should be replaced with ``What mechanism damps
the eccentricities of giant planets?''  Unless giant planets form via
gravitational instability, interactions with a gas disk are not an
option, since the eccentricities would have been excited after the gas
was cleared.  Both dynamical friction within a planetesimal disk and
planetesimal scattering could damp eccentricities in both the Solar
System and other planetary systems.  Dynamical friction alone would
not clear the small bodies, so either accretion or ejection
would be required to satisfy observational constraints (Goldreich,
Lithwick \& Sari 2004).  Planetesimal scattering provides a natural
mechanism to simultaneously damp eccentricities and remove 
small bodies from planetary systems.  

Perhaps, {\em the key parameter that determines whether a planetary
system will have eccentric or nearly circular orbits is the amount of
mass in planetesimals at the time of the last strong planet-planet
scattering event.}
This could explain why more massive planets may tend to have larger
eccentricities.  Additionally, the chaotic evolution of multiple
planet systems naturally provides a large dispersion in the time until
dynamical instability results in close encounters (Chambers, Wetherill
\& Boss 1996; FHR; Marzari \& Weidenschilling 2002; Chatterjee et al.\
2008).  This could explain why some planetary systems have large
eccentricities (late stage instability when there was little disk to
damp eccentricities), while our planets in the Solar System have
nearly circular orbits (last instability occurred while sufficient
planetesimal disk to damp eccentricities).
Unfortunately, this model would significantly complicate the
interpretation of the observed eccentricity distribution for
some extrasolar planets.  
Investigations of dynamical instabilities in systems with three or
more planets have only begun to explore the large available parameter
space.  Future numerical investigations will be necessary
to test such theoretical models.

\acknowledgements 
We thank E.I.\ Chiang, M.\ Holman, G.\ Laughlin, M.H.\ Lee,
H.\ Levison, G.W.\ Marcy, A.\ Morbidelli, J.C.B.\ Papaloizou, S.\ Peale,
and J.\ Wright for useful discussions and the anonymous referee for a thoughtful review.
Support for E.B.F.\ was provided by the University of Florida,
a Miller Research Fellowship
and by NASA through Hubble Fellowship grant
HST-HF-01195.01A awarded by the Space Telescope Science Institute,
which is operated by the Association of Universities for Research in
Astronomy, Inc., for NASA, under contract NAS 5-26555. 
This work was supported by NSF grants AST--0206182 and AST--0507727
at Northwestern University.

\clearpage

\begin{deluxetable}{rrrrrrrrrr}
\tablecolumns{10}
\tablewidth{0pc}
\tablecaption{Orbital Elements of Remaining Planet Following an Ejection}
\tablehead{
\colhead{} &  \colhead{} & \multicolumn{2}{c}{$m_1 = m_{(2)}$} & \multicolumn{2}{c}{$m_2 = m_{(2)}$} & \multicolumn{4}{c}{combined} \\ 
\colhead{$m_{(2)}/M$} & \colhead{$m_{(1)}/M$} & \colhead{$\left<a_f\right>$} & \colhead{$\sigma_{a_f}$} & \colhead{$\left<a_f\right>$} & \colhead{$\sigma_{a_f}$} & \colhead{$\left<e_f\right>$} & \colhead{$\sigma_{e_f}$} & \colhead{$\left<i_f\right>$} & \colhead{$\sigma_{i_f}$} \\ 
\multicolumn{8}{c}{} &  \colhead{($^{\circ}$)} & \colhead{($^{\circ}$)} 
}
\startdata
0.0012 & 0.0048 & 0.933 & 0.010 & 0.813 & 0.006 & 0.202 & 0.056 & 1.93 & 1.10 \\  
0.0015 & 0.0045 & 0.888 & 0.009 & 0.779 & 0.005 & 0.263 & 0.071 & 1.99 & 1.22 \\  
0.0018 & 0.0042 & 0.828 & 0.030 & 0.736 & 0.047 & 0.333 & 0.100 & 2.22 & 1.84 \\  
0.0021 & 0.0039 & 0.749 & 0.076 & 0.677 & 0.083 & 0.421 & 0.133 & 2.94 & 3.26 \\  
0.0024 & 0.0036 & 0.670 & 0.079 & 0.618 & 0.085 & 0.513 & 0.145 & 3.80 & 4.76 \\  
0.0027 & 0.0033 & 0.595 & 0.054 & 0.577 & 0.053 & 0.602 & 0.142 & 4.23 & 3.23 \\  
0.0030 & 0.0030 & 0.573 & 0.008 & 0.573 & 0.008 & 0.624 & 0.135 & 4.696 & 4.058 \\ 
\tableline
0.0003 & 0.0003 & 0.537 & 0.013 & 0.537 & 0.013 & 0.590 & 0.146 & 5.319 & 3.726 \\ 
0.0005 & 0.0005 & 0.543 & 0.015 & 0.543 & 0.015 & 0.573 & 0.149 & 5.615 & 4.283 \\ 
0.0010 & 0.0010 & 0.552 & 0.008 & 0.552 & 0.008 & 0.612 & 0.147 & 6.304 & 5.343 \\ 
0.0020 & 0.0020 & 0.565 & 0.008 & 0.565 & 0.008 & 0.616 & 0.140 & 5.204 & 4.456 \\ 
0.0030 & 0.0030 & 0.573 & 0.008 & 0.573 & 0.008 & 0.624 & 0.135 & 4.696 & 4.058 \\ 
0.0050 & 0.0050 & 0.586 & 0.009 & 0.586 & 0.009 & 0.640 & 0.123 & 3.802 & 3.583 \\ 
0.0100 & 0.0100 & 0.598 & 0.010 & 0.598 & 0.010 & 0.640 & 0.120 & 3.119 & 2.946 \\ 
\enddata
\end{deluxetable}

\clearpage


{\plotone{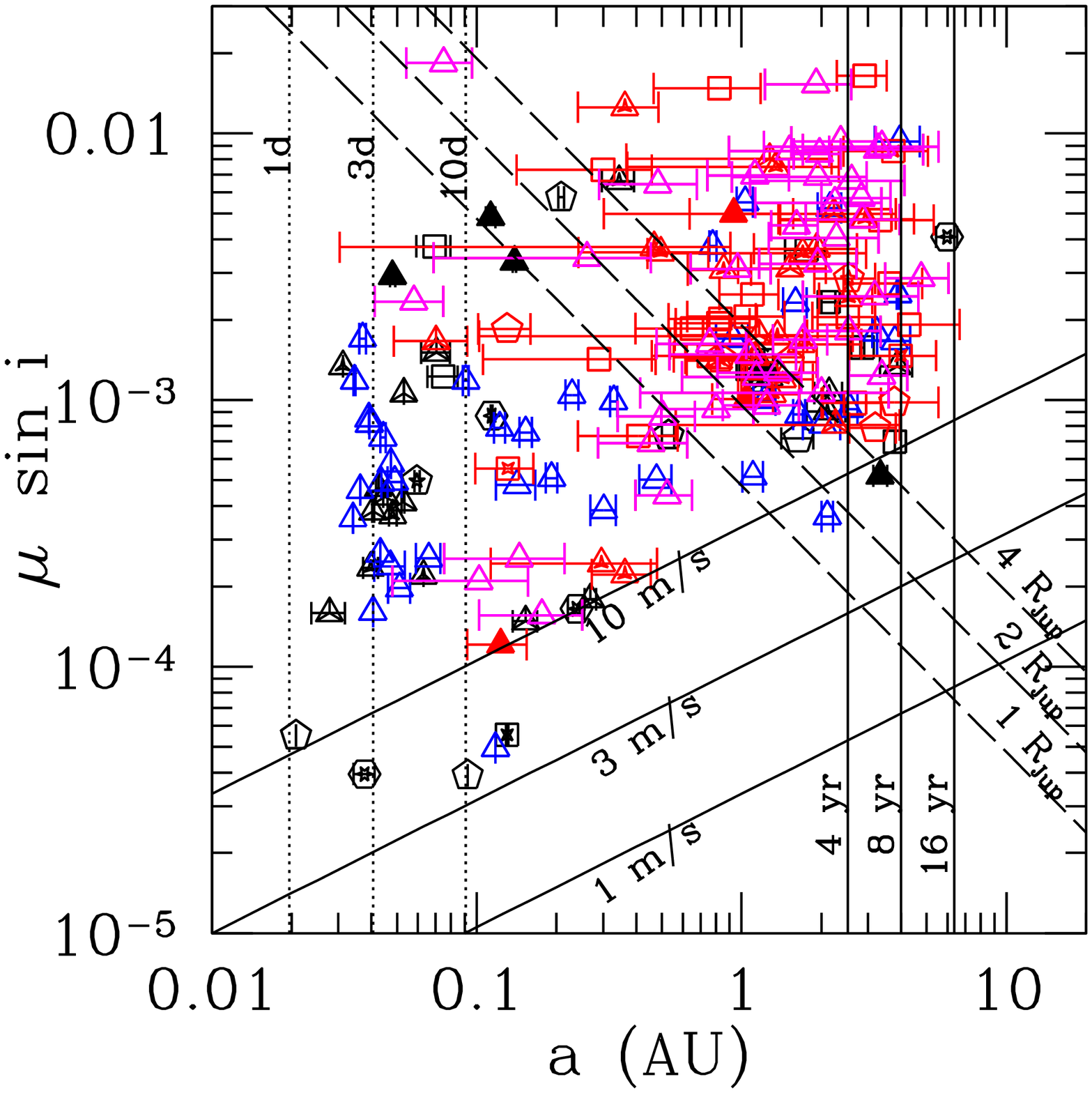}} 
\clearpage

\begin{figure}
\caption{\label{fig_avsmu}
Mass Ratio (times sine of inclination of orbital
planet to line of sight) versus semi-major axis.  Each point
corresponds to one planet discovered by radial velocity method, as
cataloged by Butler et al.\ (2006) and updated by Johnson et al.\
(2006) and Wright et al.\ (2006).  The horizontal bars on each point
indicate the periastron and apoastron distances.  The solid diagonal
lines indicate curves of constant radial velocity semi-amplitude and
roughly indicate amplitudes where radial velocity searches lose
sensitivity.  The solid vertical lines correspond to orbital periods
(assuming a stellar mass of 1$M_{\odot}$ for which the finite temporal
baseline of existing radial velocity surveys limits detection and/or
precise measurements of orbital parameters.  The dotted vertical lines
correspond to orbital periods below which tidal circularization is
likely to have damped orbital eccentricities.  The dashed diagonal
lines correspond indicate curves of constant $\theta^2\sin i\equiv \mu
a \sin i / R_p$, where $R_p$ is the planet radius and $R_{\rm Jup}$ is
the radius of Jupiter.  Planets with apoastron to the left of this
curve are not currently able to efficiently eject other bodies from
the planetary system.  Open points indicate planets around stars with
no known binary companion and no evidence for a long-term radial
velocity trend.  Points with lines extending from the center to each
vertex correspond to planets orbiting a star that also shows a long
term radial velocity trend.  Points with a star inside correspond to
planets orbiting a star in a known stellar binary (Desidera \&
Barbieri 2006).  Solid points correspond to planet orbiting stars with
both a stellar binary companion and a long term radial velocity trend.
The number of sides for each point equals to the number of known
planets orbitting the same star plus two.  Black and blue points
correspond to planets with current best-fit eccentricities less than
0.2, while red and magenta points correspond to planets with
eccentricities greater than 0.2.  Blue and magenta points correspond
to planets around stars with only one currently known extrasolar
planet, no known binary companion, and no evidence for a long-term
radial velocity trend.  Black and red points correspond to planets
orbitting stars that are known to have either two or more planets, a
binary companion, or a long-term radial velocity trend.  Magenta
points to the left of all the diagonal dashed lines correspond to
planets for which the eccentricity is unlikely to be explained by
planet-planet scattering and there is currently no evidence for
additional massive bodies orbiting their host star. 
}
\end{figure}

\clearpage

\clearpage

\clearpage

\clearpage

\clearpage

\clearpage

\clearpage


\begin{thebibliography}{}
%
\bibitem[Adams \& Laughlin(2003)]{2003Icar..163..290A} Adams, F.~C., \& Laughlin, G.\ 2003, Icarus, 163, 290 
\bibitem[Adams \& Laughlin(2006)]{2006ApJ...649.1004A} Adams, F.~C., \& Laughlin, G.\ 2006, \apj, 649, 1004 
\bibitem[Alexander et al.(2006)]{2006MNRAS.369..229A} Alexander, R.~D., 
Clarke, C.~J., \& Pringle, J.~E.\ 2006, \mnras, 369, 229 
\bibitem[]{1404} Artymowicz 1992 PASP 104, 769.
\bibitem[]{1405} Artymowicz 1993 ApJ 419, 116.
\bibitem[Barnes \& Greenberg(2006)]{2006ApJ...638..478B} Barnes, R., \& Greenberg, R.\ 2006, \apj, 638, 478 
\bibitem[Barnes \& Greenberg(2006)]{2006ApJ...647L.163B} Barnes, R., \& Greenberg, R.\ 2006, \apjl, 647, L163 
\bibitem[]{1409} Barnes, R. Qunn, T. 2004 ApJ 611, 494.
\bibitem[]{1410} Bate, M.R., Lubow, S.H., Ogilvie, G.I., Miller, K.A. 2003 MNRAS 341, 213.
\bibitem[Black(1997)]{1997ApJ...490L.171B} Black, D.~C.\ 1997, \apjl, 490, L171 
\bibitem[Bodenheimer, Laughlin, \& Lin(2003)]{2003ApJ...592..555B} Bodenheimer, P., Laughlin, G., \& Lin, D.~N.~C.\ 2003, \apj, 592, 555 
\bibitem[]{1414} Boss, A.P. 1995 Science, 267, 360.
\bibitem[Butler et al.(2006)]{2006ApJ...646..505B} Butler, R.~P., et al.\ 2006, \apj, 646, 505 
\bibitem[Chambers(1999)]{1999MNRAS.304..793C} Chambers, J.~E.\ 1999, \mnras, 304, 793 
\bibitem[]{1424} Chambers, J.E., Wetherill, G.W. \& Boss, A.P. 1996 Icarus 119, 261.
\bibitem[]{1425} Chatterjee, S., Ford, E.B., Rasio, F.A. 2008, ApJ, in press. [arxiv:astro-ph/0703166]
\bibitem[Chauvin et al.(2006)]{2006A&A...456.1165C} Chauvin, G., Lagrange, A.-M., Udry, S., Fusco, T., Galland, F., Naef, D., Beuzit, J.-L., \& Mayor, M.\ 2006, \aap, 456, 1165 

\bibitem[]{1428} Chiang, E.I. 2003 ApJ 584, 465.
\bibitem[]{1430} Chiang, E.I. \& Murray, N. 2002 ApJ 576, 473.
\bibitem[Cresswell et al.(2007)]{2007A&A...473..329C} Cresswell, P., Dirksen, G., Kley, W., \& Nelson, R.~P.\ 2007, \aap, 473, 329 
\bibitem[Cumming et al.(2008)]{2008arXiv0803.3357C} Cumming, A., Butler, 
R.~P., Marcy, G.~W., Vogt, S.~S., Wright, J.~T., \& Fischer, D.~A.\ 2008, PASP 120, 531. 
\bibitem[]{1433} D'Angelo, G., Kley, W, Henning, T. 2003 ApJ 586, 540.
\bibitem[Desidera \& Barbieri(2007)]{2006astro.ph.10623D} Desidera, S., \& Barbieri, M.\ 2007, A\&A, 462, 1039.   
\bibitem[]{1440} Faber, J.A., Rasio, F.A. \& Willems, B. 2005 Icarus, 175, 248 
\bibitem[Fischer et al.(2005)]{2005ApJ...620..481F} Fischer, D.~A., et al.\ 2005, \apj, 620, 481 
\bibitem[Ford(2005)]{2005AJ....129.1706F} Ford, E.~B.\ 2005, \aj, 129, 1706 
\bibitem[Ford(2006)]{2006ApJ...642..505F} Ford, E.~B.\ 2006, \apj, 642, 505 
Orbits of Extrasolar Planets 
\bibitem[]{1445} Ford, E.B. \& Chaing, E.I. 2007, ApJ, 661, 602.
\bibitem[]{1446} Ford, E.B. \& Gaudi, B.S. 2006, ApJ, 652, L137.
\bibitem[]{1447} Ford, E.B., Havlickova, M. \& Rasio, F.A. 2001 Icarus 150, 303.
\bibitem[]{1448} Ford, E.B. \& Holman, M. 2007, ApJL 664, 51.
\bibitem[]{1449} Ford, E.B., Kozinsky, B., Rasio, F.A. 2000 ApJ 535, 385.
\bibitem[]{1450} Ford, E.B., Lystad, V., Rasio, F.A. 2005 Nature 434, 873.
\bibitem[Ford et al.(2008)]{2008arXiv0801.2591F} Ford, E.~B., Quinn, S.~N., 
\& Veras, D.\ 2008, ApJ, in press. [arXiv:0801.2591] 
\bibitem[]{1451} Ford, E.B. \& Rasio, F.A. 2006, ApJ, 638, L45.
\bibitem[]{1452} Ford, E.B., Rasio, F.A., Yu, K. 2003 Scientific Frontiers in Research on Extrasolar Planets, eds. D. Deming \& S. Seager (ASP Conference Series, 294), 181.
\bibitem[]{1454} Gaudi, B.S. \& Winn, J.N. 2007, ApJ 655, 550.
\bibitem[]{1456} Gladman, B. 1993 Icarus, 106, 247.
\bibitem[]{1457} Goldreich, P., Lithwick, Y., Sari, R. 2004 ApJ 614, 497.
\bibitem[]{1458} Goldreich, P., Sari, R. 2003 ApJ 585, 1024.
\bibitem[]{1459} Goldreich, P., Tremaine, S. 1979 ApJ 233, 857.
\bibitem[]{1460} Goldreich, P., Tremaine, S. 1980 ApJ 241, 425.
\bibitem[Gregory(2005)]{2005ApJ...631.1198G} Gregory, P.~C.\ 2005, \apj, 631, 1198 
\bibitem[Gregory(2006)]{2006MNRAS.tmp.1375G} Gregory, P.~C.\ 2006, \mnras, 1375 
\bibitem[]{1466} Halbwachs, J.L., Mayor, M., Udry, S. 2005 A\&A 431, 1129.
\bibitem[]{1467} Holman, M., Touma, J., Tremaine, S. 1997 Nature 386, 254.
\bibitem[Hurley \& Shara(2002)]{2002ApJ...565.1251H} Hurley, J.~R., \& Shara, M.~M.\ 2002, \apj, 565, 1251 
\bibitem[Johnson et al.(2006)]{2006ApJ...647..600J} Johnson, J.~A., et al.\ 2006, \apj, 647, 600 
\bibitem[Jones et al.(2006)]{2006MNRAS.369..249J} Jones, H.~R.~A., Butler, R.~P., Tinney, C.~G., Marcy, G.~W., Carter, B.~D., Penny, A.~J., McCarthy, C., \& Bailey, J.\ 2006, \mnras, 369, 249 
\bibitem[]{1585} Juric, M. \& Tremaine, S. 2008, ApJ, in press. [arxiv:astro-ph
/0703160]
\bibitem[]{1475} Kley, W. 2000 MNRAS 313, L47.
\bibitem[]{1476} Kley, W., Lee, M.H., Murray, N., Peale, S.J. 2005 A\&A 437, 727.
\bibitem[]{1477} Kley, W., Peitz, J. \& Bryden, G. 2004 A\&A, 414, 735.
\bibitem[Kley, Lee, Murray, \& Peale(2005)]{2005A&A...437..727K} Kley, W., Lee, M.~H., Murray, N., \& Peale, S.~J.\ 2005, \aap, 437, 727 
\bibitem[]{1481} Kokubo, E. \& Ida, S. 1998a Icarus 123, 180.
\bibitem[]{1482} Kokubo, E. \& Ida, S. 1998b Icarus 131, 171.
\bibitem[Laughlin \& Adams(1998)]{1998ApJ...508L.171L} Laughlin, G., \& Adams, F.~C.\ 1998, \apjl, 508, L171 
\bibitem[]{1491} Lee, M.H. \& Peale, S.J. 2002 ApJ 567, 596.
\bibitem[]{1492} Lee, M.H. \& Peale, S.J. 2003 ApJ 592, 1201.
\bibitem[Lee et al.(2008)]{2008arXiv0801.1926L} Lee, A.~T., Thommes, E.~W., 
\& Rasio, F.~A.\ 2008, ArXiv e-prints, 801, arXiv:0801.1926 
\bibitem[]{1493} Levison, H.F., Lissuaer, J.J., Duncan, M.J. 1998 AJ 116, 1998.
\bibitem[]{1495} Lin, D.N.C. \& Ida, S. 1997 ApJ, 447, 781
\bibitem[]{1496} Lissauer, J.J. 1993 ARAA, 31, 129
\bibitem[]{1497} Lissauer, J.J. 1995 Icarus 114, 217.
\bibitem[Malmberg et al.(2007)]{2007MNRAS.377L...1M} Malmberg, D., Davies, 
M.~B., \& Chambers, J.~E.\ 2007a, \mnras, 377, L1 
\bibitem[Malmberg et al.(2007)]{2007MNRAS.378.1207M} Malmberg, D., de 
Angeli, F., Davies, M.~B., Church, R.~P., Mackey, D., 
\& Wilkinson, M.~I.\ 2007b, \mnras, 378, 1207 
\bibitem[Marzari \& Weidenschilling(2002)]{2002Icar..156..570M} Marzari, F., \& Weidenschilling, S.~J.\ 2002, Icarus, 156, 570 
\bibitem[Marzari, Weidenschilling, Barbieri, \& Granata(2005)]{2005ApJ...618..502M} Marzari, F., Weidenschilling, S.~J., Barbieri, M., \& Granata, V.\ 2005, \apj, 618, 502 
\bibitem[]{1509} Mazeh, T., Krymolowski, Y., Rosenfield, G. 1997, ApJ, 447, L103.
\bibitem[Moorhead \& Adams(2005)]{2005Icar..178..517M} Moorhead, A.~V., \& Adams, F.~C.\ 2005, Icarus, 178, 517 
\bibitem[Moorhead \& Adams(2008)]{2008Icar..193..475M} Moorhead, A.~V., \& Adams, F.~C.\ 2008, Icarus, 193, 475 
\bibitem[]{1515} Murray, N. Hansen, B., Holman, M., Tremaine, S. 1998 Science 279, 69.
\bibitem[Naef et al.(2001)]{2001A&A...375L..27N} Naef, D., et al.\ 2001, \aap, 375, L27 
\bibitem[]{1518} Nagasawa, M., Lin, D.N.C., Ida, S. 2003 ApJ 586, 1374.
\bibitem[]{1601} Nagasawa, M., Ida, S., Bessho, T. 2008, ApJ, 678, 498. 
\bibitem[Namouni \& Zhou(2006)]{2006CeMDA..95..245N} Namouni, F., \& Zhou, J.~L.\ 2006, Celestial Mechanics and Dynamical Astronomy, 95, 245 
\bibitem[Namouni(2005)]{2005AJ....130..280N} Namouni, F.\ 2005, \aj, 130, 280 
\bibitem[]{1521} Narita, N. et al.\ 2007 PASJ, 60, L1. 
\bibitem[]{1523} Ogilvie \& Lubow 2003 ApJ 587, 398.
\bibitem[Papaloizou \& Larwood(2000)]{2000MNRAS.315..823P} Papaloizou, J.~C.~B., \& Larwood, J.~D.\ 2000, \mnras, 315, 823 
\bibitem[]{1526} Papaloizou, J.C.B. \& Terquem, C. 2001 MNRAS 325, 221.
\bibitem[]{1527} Papaloizou, J.C.B. \& Terquem, C. 2002 MNRAS 332, L39.
\bibitem[]{1528} Papalouizou, J.C.B., Nelson, R.P. \& Masset, F. 2001 A\&A 366, 263.
\bibitem[]{1530} Pfahl, E.\ \& Muterspaugh, M.\ 2006 ApJ, 652, 1694.
\bibitem[]{1532} Pollack, J.B., Hubickyj, O., Bodenheimer, P., Lissauer, J.J., Podolak, M. \& Greenzweig, Y. 1996 Icarus 124, 62. 
\bibitem[Pont \& Eyer(2004)]{2004MNRAS.351..487P} Pont, F., \& Eyer, L.\ 2004, \mnras, 351, 487 

\bibitem[]{1535} Portegies Zwart, S.F.\ \& McMillan, S.L.W. 2005 ApJ 633, 141.
\bibitem[]{1536} Rasio, F.A. \& Ford, E.B. 1996 Science 274, 954.
\bibitem[]{1537} Rasio, F.A., Tout, C.A., Lubow, S.H.\ \& Livio, M. 1996 ApJ, 470, 1187.
\bibitem[Rauch \& Holman(1999)]{1999AJ....117.1087R} Rauch, K.~P., \& Holman, M.\ 1999, \aj, 117, 1087 
\bibitem[Ribas \& Miralda-Escude(2006)]{2006astro.ph..6009R} Ribas, I., \& Miralda-Escude, J.\ 2007, A\&A 464, 779. 
\bibitem[S{\'a}ndor \& Kley(2006)]{2006A&A...451L..31S} S{\'a}ndor, Z., \& Kley, W.\ 2006, \aap, 451, L31 
\bibitem[Sch{\"a}fer et al.(2004)]{2004A&A...418..325S} Sch{\"a}fer, C., Speith, R., Hipp, M., \& Kley, W.\ 2004, \aap, 418, 325 
\bibitem[]{1545} Stepinsky, T.F. \& Black, D.C. 2000, A\&A, 356, 903
\bibitem[]{1546} Stepinsky, T.F., Malhotra, R., \& Black, D.C. 2000 ApJ 545, 1004.
\bibitem[Takeda et al.(2008)]{2008arXiv0802.4088T} Takeda, G., Kita, R., 
\& Rasio, F.~A.\ 2008, submitted to ApJ, ArXiv e-prints, 802, arXiv:0802.4088 
\bibitem[Takeda \& Rasio(2005)]{2005ApJ...627.1001T} Takeda, G., \& Rasio, F.~A.\ 2005, \apj, 627, 1001 
\bibitem[Takeda \& Rasio(2006)]{2006Ap&SS.tmp...19T} Takeda, G., \& Rasio, F.~A.\ 2006, \apss, 19 
\bibitem[Thommes, Duncan, \& Levison(1999)]{1999Natur.402..635T} Thommes, E.~W., Duncan, M.~J., \& Levison, H.~F.\ 1999, \nat, 402, 635 
\bibitem[Thommes, Duncan, \& Levison(2002)]{2002AJ....123.2862T} Thommes, E.~W., Duncan, M.~J., \& Levison, H.~F.\ 2002, \aj, 123, 2862 
\bibitem[Trilling et al.(1998)]{1998ApJ...500..428T} Trilling, D.~E., Benz, W., Guillot, T., Lunine, J.~I., Hubbard, W.~B., \& Burrows, A.\ 1998, \apj, 500, 428 

\bibitem[]{1554} Tsiganis, K., Gomes, R., Morbidelli, A., Levison, H.F. 2005 Nature 435, 459.
\bibitem[]{1555} Veras, D. Armitage, P.J. 2004 Icarus 172, 349.
\bibitem[]{1556} Veras, D. Armitage, P.J. 2005 ApJ, 620, L111. 
\bibitem[]{1557} Veras, D. Armitage, P.J. 2006 ApJ, 645, 1509.
\bibitem[Vidal-Madjar et al.(2003)]{2003Natur.422..143V} Vidal-Madjar, A., Lecavelier des Etangs, A., D{\'e}sert, J.-M., Ballester, G.~E., Ferlet, R., H{\'e}brard, G., \& Mayor, M.\ 2003, \nat, 422, 143 

\bibitem[]{1563} Weidenschillingm S.J. \& Marzari, F. 1996 Nature 384, 619.
\bibitem[Wisdom \& Holman(1991)]{1991AJ....102.1528W} Wisdom, J., \& Holman, M.\ 1991, \aj, 102, 1528 
\bibitem[]{1566} Wolf, A.S., Laughlin, G., Henry, G., Fischer, D.A., Marcy, G.W., Butler, R.P., Vogt, S. 2007, ApJ, 667, 659.
\bibitem[Wright et al.(2006)]{2006astro.ph.11658W} Wright, J.~T., et al.\ 2007, ApJ 657, 533. 
\bibitem[Yu \& Tremaine(2001)]{2001AJ....121.1736Y} Yu, Q., \& Tremaine, S.\ 2001, \aj, 121, 1736 
\bibitem[]{1571} Zakamska, N.L., Tremaine, S. 2004 128, 869.

\end{thebibliography}
\end{document}